\documentclass[pre,showpacs,10pt,twocolumn,superscriptaddress,floatfix,longbibliography]{revtex4-2}

\usepackage[utf8]{inputenc}
\usepackage[T1]{fontenc}
\usepackage[colorlinks=true,linkcolor=blue,urlcolor=blue,citecolor=blue,breaklinks=true]{hyperref}

\usepackage{amsmath,amsthm,amssymb}

\usepackage{graphicx}

\usepackage{makecell}
\usepackage{booktabs}
\usepackage{multirow}

\usepackage[dvipsnames]{xcolor}

\begin{document}
\renewcommand{\acknowledgmentsname}{acknowledgements}

\title{Change of persistence in European electricity spot prices}

\author{~Leonardo~Rydin~Gorj\~ao~\href{https://orcid.org/0000-0001-5513-0580}{\includegraphics[width=3.2mm]{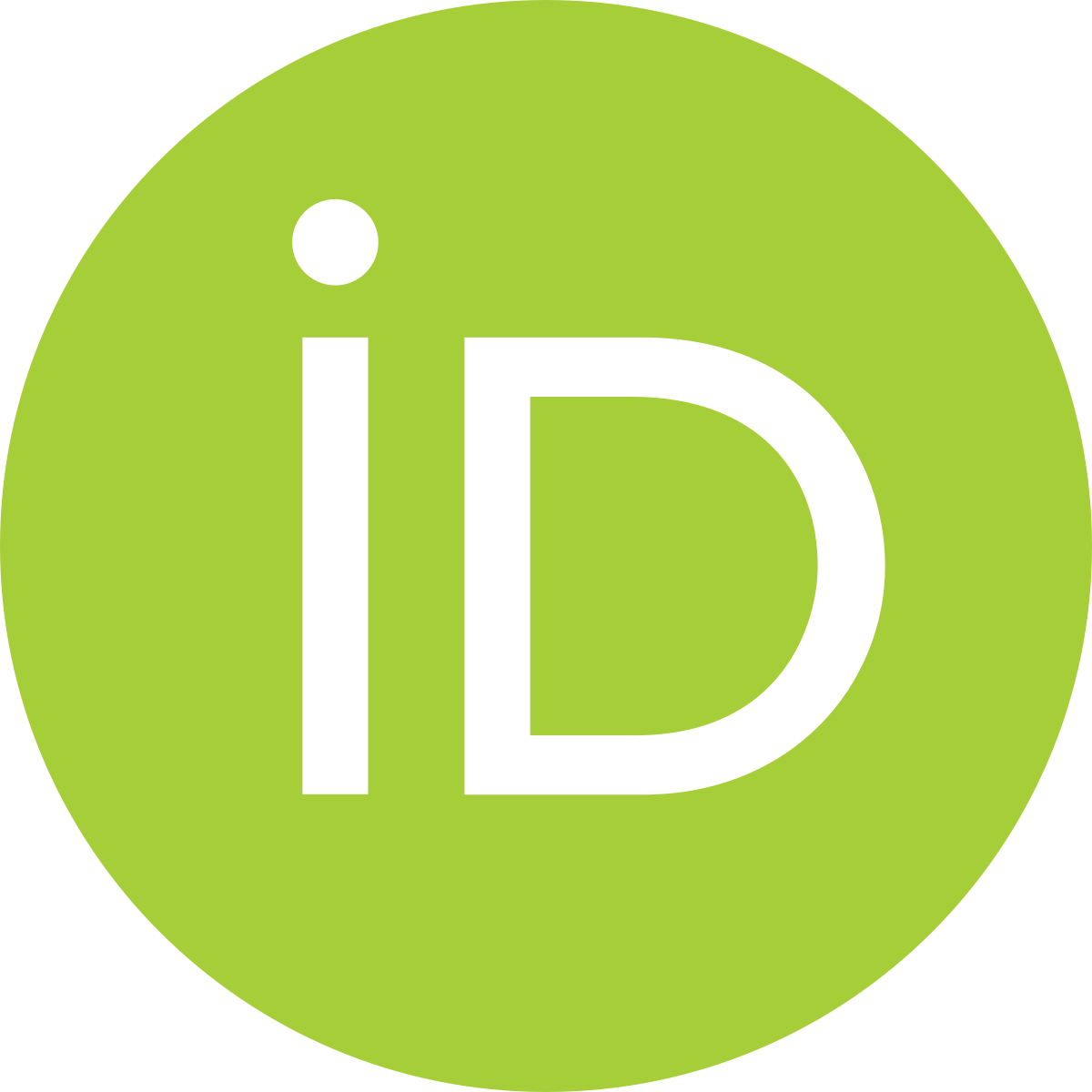}}}
\affiliation{German Aerospace Center (DLR), Institute of Networked Energy Systems, Oldenburg, Germany}
\affiliation{Department of Computer Science, OsloMet -- Oslo Metropolitan University, N-0130 Oslo, Norway}

\author{Dirk~Witthaut~\href{https://orcid.org/0000-0002-3623-5341}{\includegraphics[width=3.2mm]{orcid.png}}}
\affiliation{Forschungszentrum J\"ulich, Institute for Energy and Climate Research - Systems Analysis and Technology Evaluation (IEK-STE), 52428 J\"ulich, Germany}
\affiliation{Institute for Theoretical Physics, University of Cologne, 50937 K\"oln, Germany}

\author{Pedro~G.~Lind~\href{https://orcid.org/0000-0002-8176-666X}{\includegraphics[width=3.2mm]{orcid.png}}}
\affiliation{Department of Computer Science, OsloMet -- Oslo Metropolitan University, N-0130 Oslo, Norway}
\affiliation{NordSTAR -- Nordic Center for Sustainable and Trustworthy AI Research,
N-0166 Oslo, Norway}
\affiliation{Artificial Intelligence Lab, Oslo Metropolitan University, N-0166 Oslo, Norway}

\author{Wided~Medjroubi~\href{https://orcid.org/0000-0002-2274-4209}{\includegraphics[width=3.2mm]{orcid.png}}}
\affiliation{German Aerospace Center (DLR), Institute of Networked Energy Systems, Oldenburg, Germany}

\begin{abstract}
    The European Power Exchange has introduced day-ahead auctions and continuous trading spot markets to facilitate the insertion of renewable electricity.
    These markets are designed to balance excess or lack of power in short time periods, which leads to a large stochastic variability of the electricity prices.
    Furthermore, the different markets show different stochastic memory in their electricity price time series, which seem to be the cause for the large volatility.
    In particular, we show the antithetical temporal correlation in the intraday 15 minutes spot markets in comparison to the day-ahead hourly market.
    We contrast the results from Detrended Fluctuation Analysis (DFA) to a new method based on the Kramers--Moyal equation in scale.
    For very short term ($<12$ hours), all price time series show positive temporal correlations (Hurst exponent $H>0.5$) except for the intraday 15 minute market, which shows strong negative correlations ($H<0.5$).
    For longer term periods covering up to two days, all price time series are anti-correlated ($H<0.5$).
\end{abstract}

\maketitle

\section{Introduction}

Electricity markets have undergone far-reaching changes due to liberalisation and un-bundling of services as well as the transition to renewable energy sources. 
Short-term electricity spot markets have been established across the world~\cite{Parag2016,Gruendinger2017,Mayer2018,Koch2018,Badesa2021}, including the spot markets of the European Power Exchange (EPEX)~\cite{EPEX} and Nord Pool AS~\cite{NordPool} serving countries in Central and Northern Europe, respectively.
These markets are particularly important for the integration of non-dispatchable renewable power sources as they allow them to participate in the markets and make short-term adjustments~\cite{Edenhofer2013,Spodniak2021}.
Solar energy, for instance, if not aided with storage technologies, can only provide power during the day and generation may differ considerably from day-ahead forecasts~\cite{Braun2018,REN2021}.
Non-dispatchable power suppliers rely comparatively more (or entirely) on the ability to sell their power at an attractive price on the spot markets than long-term producers like fossil fuel and nuclear energy producers, which may rely on over-the-counter futures~\cite{Mayer2018}.
The increasing prevalence of spot markets and short-window trading has led to an increased need for accurate price models~\cite{Rodriguez2014,Conejo2018}, which can capture the full dynamics of prices at long and short time scales~\cite{Macedo2020}. 
To facilitate the development of prediction models, a more in-depth analysis of different price characteristics is necessary, in particular, to disentangle the effects of stochastic memory and extreme price events~\cite{Gruendinger2017}.

The dynamics of market prices is essentially determined by the load and supply curves, i.e. the total amount of power consumed and generated in the respective bidding zone viewed as a function of the market price~\cite{Peters2020}. 
In equilibrium, generation and load are balanced and the market price is found at the intersection point of the load and supply curve~\cite{Sensfuss2008}. 
This view can be simplified by noting that the demand is largely inelastic and the generation by variable renewable sources is essentially fixed by the weather~\cite{Infield2020}. 
Hence, market equilibrium is reached when the supply equals the difference of the demand and the variable renewable generation, commonly referred to as the residual load.
In this simplified picture, dispatchable generating units are ``switched on'' according to their variable costs until the residual load is met, which is called \textit{the merit-order effect}~\cite{Cludius2014,Kiesel2017}. 
The market price then equals the variable costs of the last generating unit, and is a monotonically increasing function of the residual load~\cite{Hirth2013,Ketterer2014,Hirth2015}.
The merit-order effect explains the general dependency of electricity prices on the load and the renewable generation, but it certainly does not cover all aspects of electricity price time series. 
A recent examination by Kremer \textit{et al.}~\cite{Kremer2020} reveals that particularly the 15-min spot market does not follow a merit-order effect during the night.
Furthermore, it cannot explain the occurrence of negative prices or extremely high prices~\cite{Cuaresma2004} and it does not capture the fluctuations around the general trend. 
In fact, any close examination of price time series unveils its strong stochastic nature~\cite{Schwartz1997,Weron2004b, Han2021}.

Electricity price dynamics shows complex behaviour, including extreme peaks, jumps, and negative values as well as persistence and long-range dependence, observed especially in the 15-min spot market prices~\cite{Weron2002,Braun2018,Kremer2021}. 
Negative electricity prices -- that emerge from an inflexible power generation and low demand -- have been increasing steadily in frequency and duration in the last 15 years~\cite{Halbruegge2021}.
Stochastic models for electricity prices can cover a large range of characteristics~\cite{Mandelbrot2004,Keles2012,Weron2014}, e.g. regime-switching~\cite{Weron2004}, jumps~\cite{Huisman2003}, seasonal components~\cite{Janczura2013,Kiesel2019}, long-range dependence~\cite{Norouzzadeh2007,AlvarezRamirez2010,Wang2013a,Wang2013b,Ergemen2016}.
These can be modelled in discrete-time, using for example auto-regressive models~\cite{Tan2010,Liu2013,Macedo2020}, or using continuous-time models~\cite{Deng2020}.

To design accurate price models, it is fundamental to understand the stochastic nature of prices, in particular the presence of memory, long-range dependence, or persistence~\cite{Grassberger1984,Halsey1986,Peng1994,Kantelhardt2002}.
The presence of temporal correlations in price time series, i.e. the non-Markovian element of prices, is a key aspect, as it dictates the statistics at very short time increments.
Correlated increments will lead to repeated dynamics in short time spans.
Conversely, anti-correlated increments will generate the opposite behaviour.
Correlations and memory effects can be quantified and examined in terms of the Hurst exponent $H$ which quantifies how the cumulative fluctuations of the increments increase with time~\cite{Hurst1951,Simonsen2003,Curpek2019}.

In this article, we will focus on price fluctuations and show that the dynamics of price fluctuations is very different for the 15-min markets as compare to all the other markets.
Particularly, we will focus on long-range dependence and extract the Hurst exponent for different price time series using both the model-independent Detrended Fluctuation Analysis (DFA) and a novel model-dependent description, which we denote Kramers--Moyal (KM) description, since it is based on the eponymous Kramers--Moyal equation for stochastic processes~\cite{vanKampen2007}.
We later connect the Hurst exponents estimated at different time scales with solar ramping effects, particularly to emphasise that the intraday 15-min markets act as balancing system to the larger, longer window markets.

In this article, market and prices will always refer to the electricity markets at EPEX and Nord Pool.
The article in organised as follows: In Sec.~\ref{sec:background} we introduce the relevant structures of spot markets.
In Sec.~\ref{sec:maths} we present the model-independent DFA and model-dependent KM description to estimate the Hurst exponent.
In Sec.~\ref{sec:results} we present an analysis centred on the EPEX spot market, with a few remarks on the Nord Pool spot market.
Lastly, in Sec.~\ref{sec:conclusion} we draw some conclusions on the changes of persistence at the different market schemes in the EPEX and Nord Pool spot markets.

\section{Background}\label{sec:background}

The electricity markets in central Europe and the Nordic countries are segmented in different time windows and contract durations~\cite{Edenhofer2013,Mayer2018}.
The over-the-counter market serves a large range of delivery periods, with electricity futures being placed from week-long up to 6 years ahead of the delivery~\cite{Hirth2013}.
Spot markets are a collection of short-term markets designed to facilitate balancing effects with the increase of renewable electricity in the grid. 
The total trading volume on the EPEX spot in 2020 was 624.8\,TWh~\cite{EPEX_record}.

The spot market is divided into two distinct schemes, the day-ahead spot market (DA) and the intraday spot market (ID)~\cite{MaerkleHuss2018}.
The DA market is organised as an auction market, where bids are placed and the market is cleared the day before delivery and consumption.
The final DA electricity prices are obtained through the intersection of demand and supply curves.
Hence, the last bid that satisfies demand sets the electricity price which is known as the \textit{market-clearing price}.
I would prefer: Since 2011, electricity can also be traded continuously on the ID market, which opens after the DA auction is cleared~\cite{EPEX_15min}.
On the ID market, electricity is traded continuously up until 5 minutes before delivery.
The continuous ID market is further divided into \textit{hourly}, \textit{30-minutes}, and \textit{15-minutes} spot markets.
This market serves as a balancing system that permits market participants to make real-time adjustments.
As part of the continuous trading on the ID markets, every contract entails a different electricity price and different prices for the same time period are possible. 
As a reference for the price distribution, the EPEX spot publishes indices of the ID price. 
The indices are the weighted averages over all trades (ID-Index), the last 3 hours prior to the delivery (ID3-Index), and the last hour prior to the delivery (ID1-Index).

The creation and use of short-window markets is directly connected with the increase in penetration of renewable energy generation. 
Particularly, solar power generation will ``ramp'' up or down at a high gradient in the morning and evening.
Contracts assuming a fixed power delivery for an hour or longer are clearly unsuited in this case. 
Furthermore, actual generation can differ considerably from day-ahead forecasts such that intra-day adjustments are necessary. 
Remarkably, in Germany, the increase of renewable energy generation was \emph{not} associated with an increase in balancing power~\cite{Hirth2015}. 
Renewables do contribute to balancing power demands, especially solar power ramps and forecast errors~\cite{Kruse2021a,Kruse2021b}, but the development of short-term ID markets and other changes in market design and system operation overcompensated this development~\cite{Ocker2017}.
  
\section{Methods}\label{sec:maths}

In this section, we will introduce the auto-correlation function, as well as the two methods employed in this study to extract the Hurst exponent, which is taken as a proxy for the persistence of time series.
The following derivations shed light on the question of discerning monofractal and multifractal behaviour, in which persistence and long-range dependence are particular aspects.

\subsection{Increment time series, structure function, and auto-covariance}

Our analysis is based on the increments of a variable $x(t)$, whose values composes a time series -- in our case, the price. 
These increments are defined as
\begin{equation}\label{eq:inc}
    \Delta x_\tau(t) = x(t+\tau)-x(t).
\end{equation}
While the increments of $x(t)$ form an additional time series for each specific value of $\tau$, their statistical moments of order $n$, $\langle \Delta x_\tau^n \rangle$, are functions of the incremental lag $\tau$.
In the context of the theory of turbulence and fluid mechanics, $\langle \Delta x_\tau^n \rangle$ is also called the structure function, and one of its main features is that it can be written as powers of the incremental lag
\begin{equation}\label{eq:scaling}
    S_n(\tau) = \langle \Delta x_\tau^n \rangle = C_n \tau^{\xi(n)} \,,
\end{equation}
with $C_n$ an amplitude (dependent on $n$) and $\xi(n)$ our function of interest, as it constitutes all the scaling and correlation properties of a time series~\cite{KoscielnyBunde1998}.
Stochastic processes like price time series can be of two natures: monofractal (linear) or multifractal (non-linear).
While multifractal processes show a non-linear $n$-dependence of $\xi(n)$, a monofractal process obeys $\langle \Delta x_\tau^n\rangle\sim\langle \Delta x_\tau^2\rangle^{n/2}$, and thus $\xi(n) = n\xi(2)/2$, showing a linear growth of the exponent $\xi(n)$ with $n$.
Thus, monofractal processes are characterised by a single/mono scaling exponent.
Multifractal processes require a multitude of exponents to be described.

One particular element of the scaling function $\xi(n)$ is $\xi(2)$, which reflects the memory of a time series, the the main focus of this work.
To first evaluate the presence of memory effects, we turn to the auto-covariance function, which provides a rough estimate of memory of a time series.
For incremental time series the auto-covariance is given by
\begin{equation}\label{eq:autocov}
    C(t-t') = \left\langle(\Delta x_\tau(t)-\mu)(\Delta x_\tau(t')-\mu)\right\rangle,
\end{equation}
with $\mu$ the mean of $\Delta x_\tau(t)$ and $\langle\cdot\rangle$ the expected value. 
If the increments are independent then $C(t-t')=0$, for $t\geq t'$.
That is to say, the correlation between any two increments is zero except if these are the same (i.e. at $t=t'$)
This is called the Markov property and processes or time series with this property are sometimes called ``memoryless''.
The auto-covariance gives a rough estimation of the memory of the time series, but in order to pin down exactly the strength of correlation or anti-correlation we need a quantitative measure of ``memory''.
For such we introduce the Hurst exponent. 

\subsection{Hurst exponent, scaling of fluctuations, and correlations}

In order to quantify exactly the strength of the correlation in an incremental time series we can examine the Hurst exponent $H\in [0,1]$ of each time series.
The Hurst exponent is a measure of the memory or long-range dependence of a process.
A process with vanishing auto-covariance has a Hurst exponent $H=0.5$.
The most common example being the uncorrelated Gaussian noise, which has no memory of the past, such that $H=0.5$.
Processes with a Hurst exponent $H>0.5$ are positively correlated increments, and conversely processes with $H<0.5$ are negatively correlated increments.
In relation to the scaling function in Eq.~\eqref{eq:scaling}, the Hurst exponent is $H=\xi(2)/2$~\cite{Tabar2019}.

At this point, it is important to note that a time series can have various Hurst exponents at different time scales.
As we have introduced, once we deal with increment statistics we move from a picture in time $t$ to a picture in scale $\tau$.
This said, a process can have different Hurst exponents $H$ at different time scales $\tau$.
For example, a price time series may have $H>0.5$, i.e. positively correlated increments, at the shortest scale of $\tau=1$ hour, but $H<0.5$, i.e. negatively correlated increments, at $t=24$ hours. 
This would imply that the process has two very distinct behaviours at different time steps.
On the time scale of few hours ($\tau=1$ hour), the process tries to repeat itself. 
This would mean that for price time series the shortest scale of $\tau=1$ hour we find $H>0.5$, i.e. positively correlated processes, but at $t=24$ hours the Hurst exponent is $H<0.5$, i.e. negatively correlated processes, implying the process has two very distinct behaviours: in the time scales of few hours ($\tau=1$ hour), the process tries to repeat itself. 
If the price is going down, it will likely continue to go down, and if it is going up, it will likely continue to go up.
On the opposite side, in the timescale of days ($\tau=24$ hours), the process will try to move in the opposing manner. 
If prices were going up, these will likely go down and vice versa.
It is paramount to note that this information is central to understanding if hedging or arbitraging is conceptually possible in these markets.
Knowing the Hurst exponent indicates, statistically, what is the likelihood that a process repeats or opposes itself in different time scales.
We remind the reader again that here we are focused on the fluctuations of the prices, thus not the price itself, which generally follows the residual load.

In order to estimate the Hurst exponent $H$ from time series, we now introduce two methods employed in this article.

\subsection{Multifractal detrended fluctuation analysis}

Multifractal detrended fluctuation analysis (MFDFA)~\cite{Kantelhardt2002,Ihlen2012,RydinGorjao2021} is one of the principal methods to estimate the scaling function $\xi(n)$ of one-dimensional time series.
It is well suited for time series with trends, like price time series.
It generalises detrended fluctuation analysis~\cite{Peng1994,Peng1995}, which can only estimate monofractal properties of time series.
In here we only include a brief outline of the MFDFA algorithm, focusing on the properties it unveils, not the actual numerical procedure.
The algorithm's estimation procedure can be found in the original publication by Kantelhardt \textit{et al.}~\cite{Kantelhardt2002} and a more descriptive and detailed examination of the algorithm is found in Ref.~\cite{Ihlen2012}.

The MFDFA algorithm seeks to unveil the scaling of the increments in Eq.~\eqref{eq:scaling} by segmenting the data into small snippets of a given size $\tau$, $\tau$ here being the same as the incremental lag.
(1) To examine the fluctuations ``around'' the general trend of the data, polynomials are subtracted to the data separately for each snippet.
(2) Then the variance of the detrended price time series is calculated for each snippet and the result is averaged over all snippets of the time series.
(3) Finally, the procedure is repeated for increasing snippet sizes $\tau$, which allows one to study the increase in the average variance of the snippets as $\tau$ increases.
This is none other that the function $\xi(2)$ in Eq.~\eqref{eq:scaling}.
This procedure is denoted DFA$m$, where $m$ indicated the choice of order of the aforementioned polynomial fits.
The MFDFA includes a variation on this procedure, comprising an additional layer by taking a set of powers whilst averaging the variances.
This allows the study of the multifractal structure of the process, i.e. whether $\xi(n)$ is a linear or non-linear function of $n$.  

\subsection{The Kramers--Moyal description of incremental statistics}

The Kramers--Moyal (KM) description of incremental time series, which we introduce now, is a novel method in the context of deriving scaling functions, which relies on the description of the increments of the stochastic variable.
While it is unable to deal with non-stationary series, this method is computationally far less demanding than MFDFA.

The KM description represents the evolution of the probability density $\rho(\Delta x_\tau,\tau)$ of the increments of a time series obeying the Kramers--Moyal equation~\cite{Kramers1940,Moyal1949,Friedrich1997,Mueller2018,Friedrich2018,Tabar2019,Friedrich2020}
\begin{equation}\label{eq:KM}
    -\frac{\partial \rho(\Delta x_\tau,\tau)}{\partial \tau} = \sum\limits_{M=1}^\infty (-1)^m\frac{\partial^m }{\partial \Delta x_\tau^m}\bigg[ D_n(\Delta x_\tau,\tau) \rho(\Delta x_\tau,\tau)\bigg],
\end{equation}
where the Kramers--Moyal coefficients are given by
\begin{equation}
\begin{aligned}
   D_m&(\Delta x_\tau,\tau) = \\&\frac{1}{m!} \lim\limits_{s\to \tau}\frac{1}{s-\tau}\int (\Delta x_s - \Delta x_\tau)^m \rho(\Delta x_\tau,\tau|\Delta x_s,s) \mathrm{d}\Delta x_s.
\end{aligned}
\end{equation}
Important to notice is the independent variable of the partial differential equation in Eq.~\eqref{eq:KM}.
The independent variable of the Kramers--Moyal equation is the incremental lag $\tau$, instead of the usual time, which can be interpreted as a \textit{time scale}. 
Consequently, the Kramers--Moyal equation \eqref{eq:KM} describes the evolution of price increments (fluctuations) $\Delta x_\tau$ through ``consecutive'' time scales.
The evolution of the probability density $\rho(\Delta x_\tau,\tau)$ follows from large to small time scales, since the derivative with respect to $\tau$ is affected by a minus sign.

From here, one obtains a differential relation between the moments $\langle \Delta x_\tau^n \rangle$ by multiplying these onto Eq.~\eqref{eq:KM} and integrating with respect to $\Delta x_\tau$,resulting in~\cite{Nickelsen2017,Friedrich2018}
\begin{equation}\label{eq:moments}
   -\frac{\partial \langle \Delta x_\tau^n \rangle}{\partial \tau} = \sum\limits_{k=1}^n \frac{n!}{(n-k)!}\left\langle  \Delta x_\tau^{n-k} D_n(\Delta x_\tau,\tau)\right\rangle.
\end{equation}
For a full derivation, see Refs.~\cite{Friedrich2018,Tabar2019}.

If all higher-order Kramers--Moyal coefficients vanish ($D_n(\Delta x_\tau,\tau)=0, n>2$), i.e.~the process is described by the Fokker--Planck equation, the increments of a stochastic time series detail the correlation structure and multifractal properties, given by
\begin{equation}\label{eq:D1D2}
   D_1(\Delta x_\tau,\tau) = -H\frac{\Delta x_\tau}{\tau},\quad D_2(\Delta x_\tau,\tau) = b\frac{\Delta x_\tau^2}{\tau}
\end{equation}
with $b$ a non-negative constant.
Inserting Eq.~\eqref{eq:D1D2} into Eq.~\eqref{eq:moments} yields
\begin{equation}
   \frac{\partial \langle \Delta x_\tau^n \rangle}{\partial \tau} = \left[nH - bn(n-1) \right]\frac{\langle \Delta x_\tau^n \rangle}{\tau},
\end{equation}
which by dividing by $\langle \Delta x_\tau^n \rangle$ and discretising the derivative $\partial \ln(\langle \Delta x_\tau^n) \rangle/\partial \tau$ results in
\begin{equation}
  \xi(n) = nH - bn(n-1).
\end{equation}
If the incremental process is described more generally by the full Kramers--Moyal equation, with all Kramers--Moyal coefficient $D_n(\Delta x_\tau,\tau)$ being potentially non-vanishing,
the relation between $D_n$ and $\xi(n)$ becomes more complex, but $D_1$ and $H$ maintain their relation in Eq.~\eqref{eq:D1D2}.
That is, the Hurst exponent $H$ can be obtained unequivocally, regardless of the mono or multifractal nature of the time series.
A monofractal process will have $b=0$ and the second Kramers--Moyal coefficient will be constant (not quadratic).

\subsection{Implementation}

In order to calculate the Hurst exponent $H$ we employ the two aforementioned methods.
Regardless of the fractality of the process we can use DFA (MFDFA with $q=2$) to estimate the Hurst exponent $H$.
Utilising the KM description, we recover the Hurst exponent $H$ by estimating the first Kramers--Moyal coefficient of the increments, also known as drift.
We do this by utilising a Nadaraya--Watson estimator with an Epanechnikov kernel~\cite{RydinGorjao2019}.
In principle, the Kramers--Moyal coefficients are defined in the limit $\tau\to 0$.
In practical applications the resolution is limited, such that we resort to the smallest increments in the data set~\cite{Lamouroux2009,RydinGorjao2019,RydinGorjao2021b}.

Secondly, the MFDFA algorithm, although well adapted to long-term non-stationarity, suffers some drawbacks for periodic signals, as price time series (cf. Ref.~\cite{Movahed2006}).
For this matter, the hourly scales are obtained with DFA1 (1\textsuperscript{st}-order polynomial), and the daily scales are obtained with DFA2 (2\textsuperscript{nd}-order polynomial).
To obtain some statistics of the range of Hurst exponents using MFDFA, we estimate the Hurst exponent by fitting the $\xi(2)$ function for various ranges of the snippets within the daily and hourly scale.
In this way, the results are presented in a box-and-whiskers plot to best quantify the estimation of the Hurst exponents $H$~\cite{RydinGorjao2021}.

\section{Persistence at the hourly and daily scales}\label{sec:results}

In Europe, electricity prices generally follow the residual load and thus show a pronounced daily pattern: Electricity is consistently cheaper during the night and more expensive during the day.~\cite{Braun2018}.
However, our analysis shows that price time series paint a far more complex picture of what happens in the markets.
First, unlike many other commodities, electricity prices can become negative.
Secondly, electricity prices can display very short burst, where the price can increase by one order of magnitude~\cite{Han2021}.

When carefully examining electricity price time series, a couple of phenomena become clear.
Firstly, the distribution of prices follows very leptokurtic distributions (i.e. distributions with large tails). 
Secondly, prices exhibit strong multifractal behaviour, i.e. large jumps or long ``plateaus''.
This is a characteristic behaviour for multifractal time series, where occasionally the fluctuations of the prices become very small or very large for a short amount of time.
Lastly, the price time series show long-range dependence, i.e. memory effects at very short time scales. 
We will address this last point here and study the changing behaviour of correlations in price time series in two distinct scales: hourly scale ($<12$ hours) and daily scale ($12$ to $48$ hours).
We point the reader to Han \textit{et al.}~\cite{Han2021} for a discussion on the distribution and multifractality of price time series.
Price extrema can be very large, but are rare events.
For the day-to-day operation, understanding the stochastic memory of price development is key~\cite{Halbruegge2021}.

To examine the long-range dependence or persistence in stochastic time series one can examine the time series in their smallest scales.
Let $x(t)$ be a price time series in time, where in our case $t$ is discrete, i.e. $x(t)$ takes values every hour or every 15 minutes, depending on the market.
To distance ourselves from the daily changes and examine solely the long-range dependence in price time series, we can examine the increments of prices $\Delta x_\tau(t)$, as given in Eq.~\eqref{eq:inc}.
This new process -- and in that manner, a new time series -- entails only the elements of the stochastic fluctuations of the prices.
More importantly, it reflects both the memory of the system, i.e. its persistence or long-range dependence, as well as its fractal or multifractal structure.
To pinpoint the notion of long-range dependence, recall that \textit{prima facie} one often considers in modelling the use of Gaussian white noise, which critically has independent increments, i.e. no memory.
This is easily quantifiable, by studying for example the auto-covariance function in Eq.~\eqref{eq:autocov}.
Naturally the question is whether price time series have non-vanishing correlations, i.e. are they Markov processes?

\begin{figure}[t]
	\includegraphics[width=\linewidth]{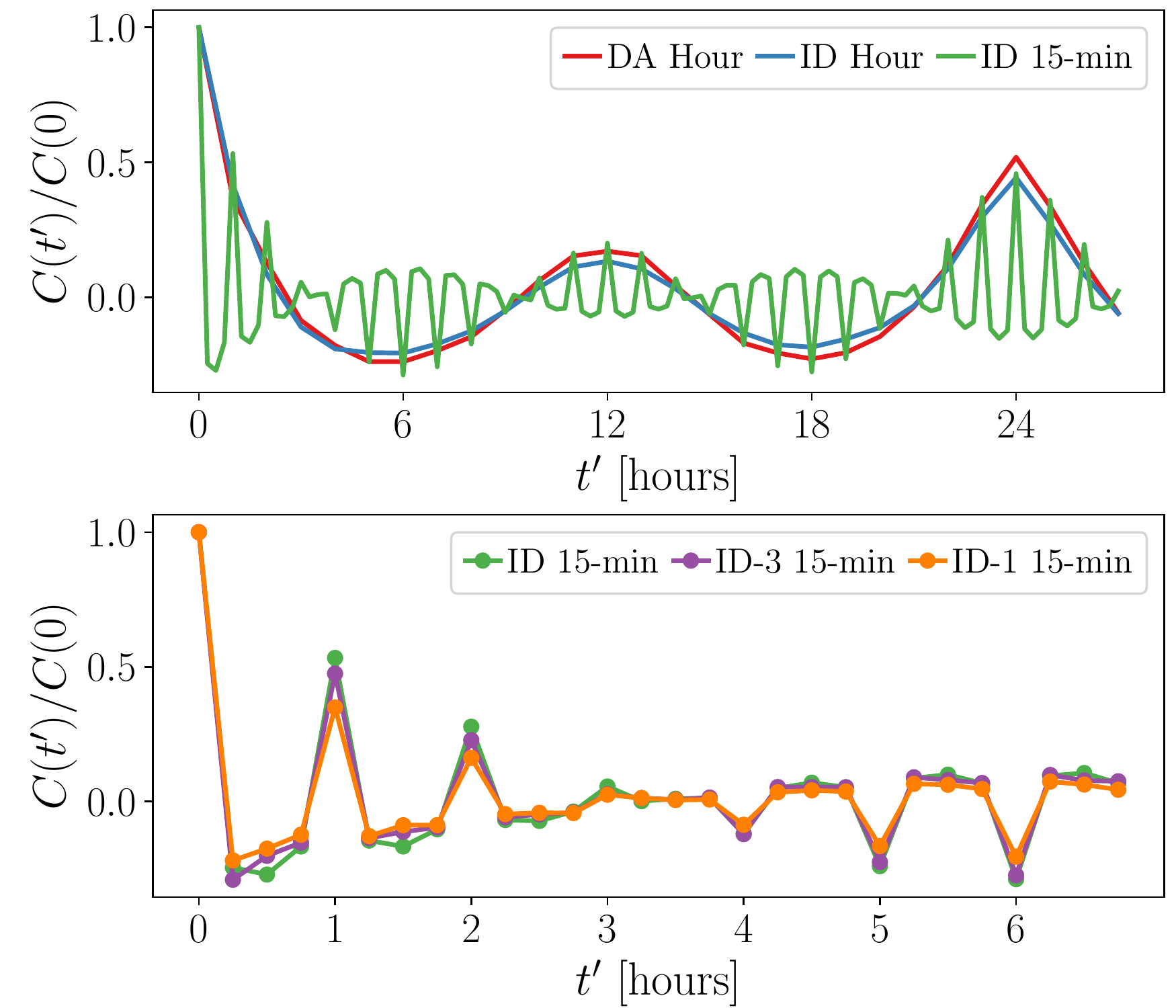}
	\caption{Normalised auto-covariance of the price increments $\Delta x_{\tau}$a t the shortest incremental lag. 
	(Top) the DA hourly, ID hourly, and ID 15-min prices,  (bottom) ID 15-min, ID-3 15-min, and ID-1 15-min prices.
	The hourly prices show positive correlations of the increments in the first 2 to 3 hours, where all ID 15-min prices show anti-correlated behaviour with the auto-covariance dropping from 1 at $t'=0$ to negative values at 15, 30, and 45 minutes.}\label{fig:1}
\end{figure}

In Fig.~\ref{fig:1} we display the the normalised auto-covariance of the price increments $\Delta x_\tau(t)$ at their smallest incremental lag $\tau=1$ (1 hour for hourly markets, 15 minutes for 15-min markets). 
The auto-covariance function is noticeably different from zero -- the increment do not have the Markov property.
In fact, it shows a very distinct behaviour for the hourly market: we see positive correlation in the first few time-lags $t'$, since the auto-covariance function is positive.
In contrast, all 15-min markets drop from $1$ at $t'=0$ to $\approx-0.25$ at $t'=15$ minutes, indicating anti-correlation.
Moreover, we note that at every hour the auto-covariance function of the 15-min market mimics the other markets.
This is our first indication that price time series are not memoryless processes.
Yet, there is a clear distinction between the memory of the hourly markets and the 15-min markets.

\begin{figure}[t]
	\includegraphics[width=\linewidth]{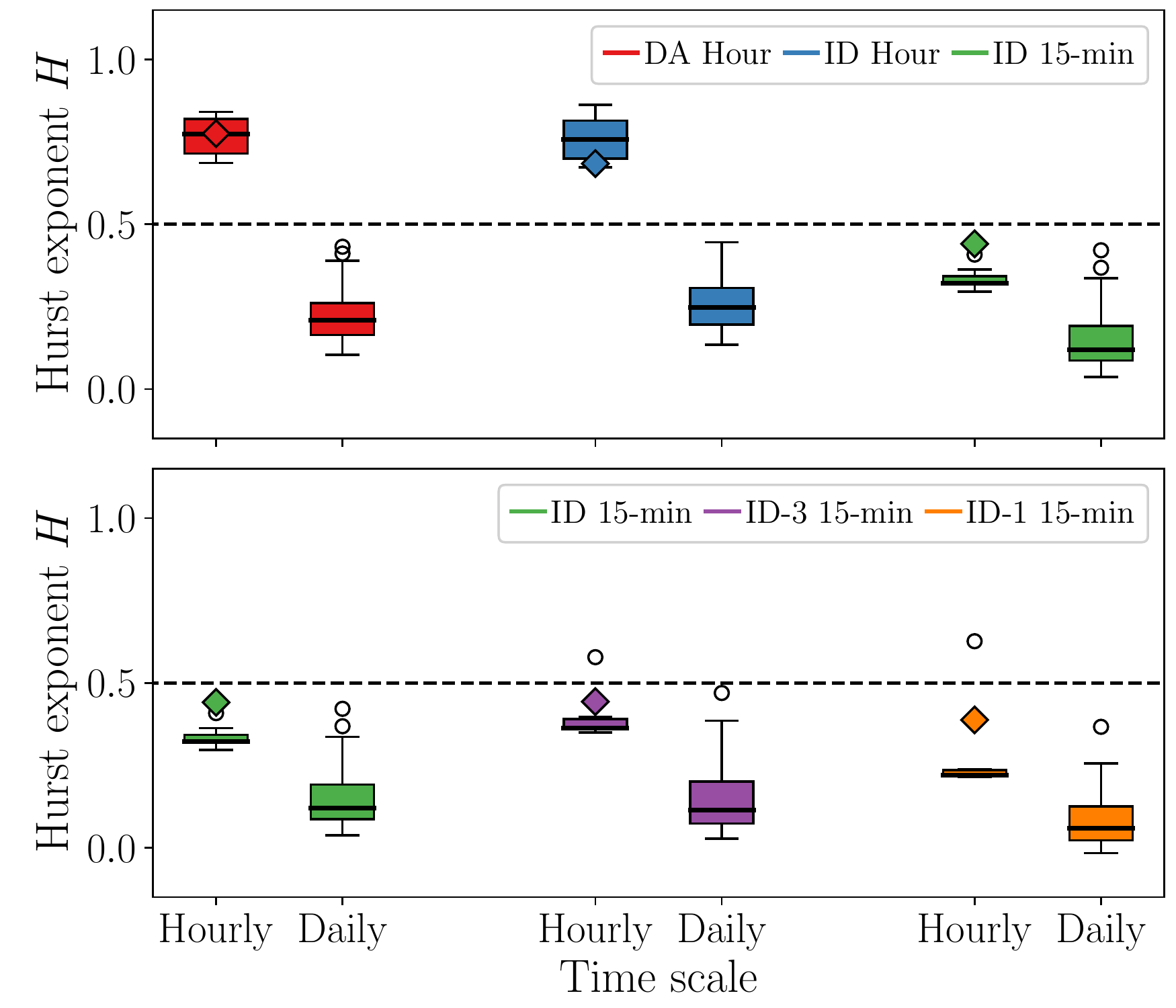}
	\caption{Hurst exponents $H$ of (top) the DA hourly, ID hourly prices, and ID 15-min prices, as well as (bottom), the ID 15-min, ID-3 15-min, and ID-1 15-min prices, all for hourly scales ($<12\,$h) and daily scales ($12\,$h to $48\,$h).
	The box-whiskers plots are obtained via MFDFA.
	The hourly scales use DFA1 (1\textsuperscript{st}-order polynomials) and the daily scales use DFA2 (2\textsuperscript{nd}-order polynomials).
    The diamond-shaped markers indicate the Hurst exponents calculated using the KM description of increments, which is only possible for small incremental lags.}\label{fig:2}
\end{figure}

In Fig.~\ref{fig:2} we display the Hurst exponent at the hourly scales ($<12$ hours), and daily scales ($12$ to $48$ hours).
We include ID-3 and ID-1 market indices, which follow an identical statistics as the full ID 15-min index, for comparison.
In the daily time scales, larger than $12$ hours, all market show anti-persistent behaviour ($H<0.5$), and all markets are identical in terms of their memory.
This in practice means that, if the prices showed a positive trend in their fluctuations, in a period of 12 to 48 hours after they will likely show negative trends.

On the hourly time scale, i.e.~for increments $\tau \le 12$ hours, the hourly and 15-min markets become distinct in their nature: where the 15-min always shows anti-persistent behaviour at all scales, at the hourly scale the hourly markets become positively correlated ($H>0.5$).
This shows that price fluctuations of the hourly markets show a repeating pattern: if one observes a price increment above average in one hour, it will likely remain above average in the following hour.

Conversely follows, if the price increment are below their average in a given hour, they are likely to stay below average in the subsequent hours.
In contrast, the 15-min market is anti-persistent, thus a price increment above average at a given time is likely followed by a price increment below average the moment after, giving rise to a characteristic jigsaw-like behaviour around the average price.

This is a direct effect of solar ramps during the day: from sunrise to it peak, solar power generation increases.
Within a one-hour trading block, the solar generation in the first 15 minutes of the hour cannot match consumption, making the ID 15-min prices higher than the average set by the hourly market.
As the hour advances, solar generation increases.
At the last 15-minute block, solar generation is in excess in comparison to the consumption, making the 15-min prices lower than the hourly average.
The opposite effect follows from the solar peak to dusk, where there is an excess of solar power generation at the first 15 minutes of the hours, and a lack thereof in the last 15 minutes.
This is the manifest difference of the ID 15-min market in comparison with the hourly markets.
We note here that this covers solely the effects during the day, where generation/consumption are high.
As pointed out in Ref.~\cite{Kremer2020}, the price dynamics at night is inextricably coupled with wind-power forecast errors.
Where the DA market is use to guarantee the delivery of power, the ID markets serve as a balancing for inapt forecasting of wind generation.

\begin{table}[t]
  \centering
  \begin{tabular}{|c|r|c|c|c|}\hline
 \multicolumn{2}{|c|}{~}  & \multicolumn{2}{c|}{Hourly scale} & Daily scale\\
 \multicolumn{2}{|c|}{~}  & $H_{\mathrm{MFDFA}}$ & $H_{\mathrm{KM}}$ & $H_{\mathrm{MFDFA}}$\\
\hline
\parbox[t]{2mm}{\multirow{5}{*}{\rotatebox[origin=c]{90}{EPEX}}} & DA Hour & 0.767 $\pm$ 0.058 & 0.775 & 0.247 $\pm$ 0.076 \\
 & ID Hour & 0.760  $\pm$ 0.068 & 0.684 & 0.280  $\pm$ 0.070  \\
\cline{2-5}
 & ID 15-min & 0.335 $\pm$ 0.035 & 0.441 & 0.179 $\pm$ 0.096 \\
 & ID-3 15-min & 0.398 $\pm$ 0.075 & 0.443 & 0.182 $\pm$ 0.111 \\
 & ID-1 15-min & 0.281 $\pm$ 0.141 & 0.388 & 0.111 $\pm$ 0.090  \\
\hline
\parbox[t]{2mm}{\multirow{3}{*}{\rotatebox[origin=c]{90}{Nord Pool}}} & SYS DA Hour & 0.770  $\pm$ 0.068 & 0.775 & 0.262 $\pm$ 0.062 \\
 & DK DA Hour & 0.707 $\pm$ 0.049 & 0.826 & 0.235 $\pm$ 0.073 \\
\cline{2-5}
 & LT DA Hour & 0.561 $\pm$ 0.043 & 0.927 & 0.421 $\pm$ 0.052 \\ \hline
\end{tabular}
\caption{Hurst exponents for the EPEX spot markets (DA hourly, ID hourly, and all 15-min ID indices) and several Nord Pool spot markets (system price (SYS), Danish price (DK), and Lithuanian price (LT)).
LT DA hourly shows a discrepancy of $H_{\mathrm{KM}}$ and $H_{\mathrm{MFDFA}}$.}\label{tab:1}
\end{table}

For comparison, we include an identical analysis for the DA hourly markets in Nord Pool.
Demonstratively, we include the system price (SYS), the Danish price (DK), and the Lithuanian price (LT), which is part of the Baltic power grid, but has its exchange within Nord Pool.
The estimation of the Hurst exponent at the hourly and daily scales are given in Tab.~\ref{tab:1}. 
The results are in line with what we have seen thus far.
The Hurst exponents $H$ show positive correlation ($H>0.5$) at the hourly scale and anti-correlation ($H<0.5$) at the daily scale for all DA hourly markets.
The weakest memory effects, i.e. the market closes to having no memory is the Lithuanian, yet interestingly, this is the one where $H_{\mathrm{KM}}$ seems to differ strongly from the MFDFA results $H_{\mathrm{MFDFA}}$.
The discrepancy is of unknown cause and could be related to complex multifractal behaviour of the prices, which hinder an accurate estimation of $H$.

\section{Conclusion}\label{sec:conclusion}

We addressed the fluctuations of price time series in the EPEX spot markets, with a particular focus on the intraday (ID) market and its 15-min trading window, in comparison with the day-ahead (DA) hourly market.
We showed that the stochastic memory in price time series in the 15-min window ID market differs strongly from the memory in the hourly markets, both in the DA hourly and ID hourly markets.
We utilised both in a model-free approach (using Detrended Fluctuation Analysis) and model-based approach (using Kramers--Moyal equation in scale).
The ID 15-min market shows a strong anti-correlation pattern in its increments ($H<0.5$) both the hourly and the daily time scale. 
This falls in line with previous observations which showed that, as solar power ramps up at the beginning of the day, it creates a jigsaw-like price phenomenon: for the first 15 minutes of the hour, generation falls short of consumption, making prices higher than average.
Conversely, at the last 15 minutes of the hour, solar generation outpaces consumption, and prices lower below the average~\cite{Kiesel2017}.
Subsequently, from the peak solar hour onward, solar generation is consistently larger than consumption in the first 15 minutes of the hours, and smaller in the last.
This inverts the jigsaw, making ID 15-min prices lower than average in the first 15 minutes, and higher than the average in the last 15 minutes.
In contrast the hourly markets show a pronounced change in this memory behaviour from correlations at the hourly scale to anti-correlations at the daily scale.

Our examination stresses the need to study and model the ID 15-min prices in a different manner from the longer-term markets prices.
Kremers \textit{et al.}~\cite{Kremer2020} points out that ID 15-min price dynamics at night are mainly driven by electricity prices themselves and wind power forecast errors.
Thus, unveiling the memory structures within price time series helps to understand the price dynamics at very short time scales. 
This is particularly relevant for solar and wind energy producers or vendors, which are highly active on short-term spot markets.
Moreover, understanding the short-term dynamics helps to design stochastic prediction models which do not rely on present knowledge of the current power generation, but can actively forecast prices knowing only the price trend.
This is important as power grid and energy systems transition into smart-grids, where access to large datasets of prices, production, and weather forecasts are present.
As these future models are yet to be implemented, producers only have access to single price time series and small datasets, and thus rely heavily on low-scale stochastic models to predict future electricity prices.

\begin{acknowledgements}
We thank Eike Cramer for valuable discussions on the pricing principles of the EPEX spot market.
This work was supported by the Helmholtz Information \& Data Science Academy (HIDA) via the \textit{HIDA Trainee Network Postdoc Scholarship} and the Helmholtz Association via the grant \textit{Uncertainty Quantification -- From Data to Reliable Knowledge (UQ)} with grant no.~ZT-I-0029.
L. R. G. and P. G. L. thank the OsloMet Artificial Intelligence Lab (Norway) and the Nordic Center for Sustainable and Trustworthy AI Research (NordSTAR) for partial financial support.

\end{acknowledgements}

\bibstyle{apsrev4-2}
\bibliography{bib}

\begin{thebibliography}{75}%
\makeatletter
\providecommand \@ifxundefined [1]{%
 \@ifx{#1\undefined}
}%
\providecommand \@ifnum [1]{%
 \ifnum #1\expandafter \@firstoftwo
 \else \expandafter \@secondoftwo
 \fi
}%
\providecommand \@ifx [1]{%
 \ifx #1\expandafter \@firstoftwo
 \else \expandafter \@secondoftwo
 \fi
}%
\providecommand \natexlab [1]{#1}%
\providecommand \enquote  [1]{``#1''}%
\providecommand \bibnamefont  [1]{#1}%
\providecommand \bibfnamefont [1]{#1}%
\providecommand \citenamefont [1]{#1}%
\providecommand \href@noop [0]{\@secondoftwo}%
\providecommand \href [0]{\begingroup \@sanitize@url \@href}%
\providecommand \@href[1]{\@@startlink{#1}\@@href}%
\providecommand \@@href[1]{\endgroup#1\@@endlink}%
\providecommand \@sanitize@url [0]{\catcode `\\12\catcode `\$12\catcode
  `\&12\catcode `\#12\catcode `\^12\catcode `\_12\catcode `\%12\relax}%
\providecommand \@@startlink[1]{}%
\providecommand \@@endlink[0]{}%
\providecommand \url  [0]{\begingroup\@sanitize@url \@url }%
\providecommand \@url [1]{\endgroup\@href {#1}{\urlprefix }}%
\providecommand \urlprefix  [0]{URL }%
\providecommand \Eprint [0]{\href }%
\providecommand \doibase [0]{https://doi.org/}%
\providecommand \selectlanguage [0]{\@gobble}%
\providecommand \bibinfo  [0]{\@secondoftwo}%
\providecommand \bibfield  [0]{\@secondoftwo}%
\providecommand \translation [1]{[#1]}%
\providecommand \BibitemOpen [0]{}%
\providecommand \bibitemStop [0]{}%
\providecommand \bibitemNoStop [0]{.\EOS\space}%
\providecommand \EOS [0]{\spacefactor3000\relax}%
\providecommand \BibitemShut  [1]{\csname bibitem#1\endcsname}%
\let\auto@bib@innerbib\@empty
\bibitem [{\citenamefont {Parag}\ and\ \citenamefont
  {Sovacool}(2016)}]{Parag2016}%
  \BibitemOpen
  \bibfield  {author} {\bibinfo {author} {\bibfnamefont {Y.}~\bibnamefont
  {Parag}}\ and\ \bibinfo {author} {\bibfnamefont {B.~K.}\ \bibnamefont
  {Sovacool}},\ }\bibfield  {title} {\bibinfo {title} {Electricity market
  design for the prosumer era},\ }\href
  {https://doi.org/10.1038/nenergy.2016.32} {\bibfield  {journal} {\bibinfo
  {journal} {Nature Energy}\ }\textbf {\bibinfo {volume} {1}},\ \bibinfo
  {pages} {16032} (\bibinfo {year} {2016})}\BibitemShut {NoStop}%
\bibitem [{\citenamefont {Gründinger}(2017)}]{Gruendinger2017}%
  \BibitemOpen
  \bibfield  {author} {\bibinfo {author} {\bibfnamefont {W.}~\bibnamefont
  {Gründinger}},\ }\href {https://doi.org/10.1007/978-3-658-17691-4} {\emph
  {\bibinfo {title} {Drivers of Energy Transition}}},\ \bibinfo {edition}
  {online}\ ed.\ (\bibinfo  {publisher} {Springer Fachmedien Wiesbaden},\
  \bibinfo {year} {2017})\ \bibinfo {note} {{ISBN}:
  978-3-658-17690-7}\BibitemShut {NoStop}%
\bibitem [{\citenamefont {Mayer}\ and\ \citenamefont
  {Tr{\"u}ck}(2018)}]{Mayer2018}%
  \BibitemOpen
  \bibfield  {author} {\bibinfo {author} {\bibfnamefont {K.}~\bibnamefont
  {Mayer}}\ and\ \bibinfo {author} {\bibfnamefont {S.}~\bibnamefont
  {Tr{\"u}ck}},\ }\bibfield  {title} {\bibinfo {title} {Electricity markets
  around the world},\ }\href {https://doi.org/10.1016/j.jcomm.2018.02.001}
  {\bibfield  {journal} {\bibinfo  {journal} {Journal of Commodity Markets}\
  }\textbf {\bibinfo {volume} {9}},\ \bibinfo {pages} {77} (\bibinfo {year}
  {2018})}\BibitemShut {NoStop}%
\bibitem [{\citenamefont {Koch}\ and\ \citenamefont {Hirth}(2019)}]{Koch2018}%
  \BibitemOpen
  \bibfield  {author} {\bibinfo {author} {\bibfnamefont {C.}~\bibnamefont
  {Koch}}\ and\ \bibinfo {author} {\bibfnamefont {L.}~\bibnamefont {Hirth}},\
  }\bibfield  {title} {\bibinfo {title} {Short-term electricity trading for
  system balancing: {A}n empirical analysis of the role of intraday trading in
  balancing {G}ermany's electricity system},\ }\href
  {https://doi.org/10.1016/j.rser.2019.109275} {\bibfield  {journal} {\bibinfo
  {journal} {Renewable and Sustainable Energy Reviews}\ }\textbf {\bibinfo
  {volume} {113}},\ \bibinfo {pages} {109275} (\bibinfo {year}
  {2019})}\BibitemShut {NoStop}%
\bibitem [{\citenamefont {Badesa}\ \emph {et~al.}(2021)\citenamefont {Badesa},
  \citenamefont {Strbac}, \citenamefont {Magill},\ and\ \citenamefont
  {Stojkovska}}]{Badesa2021}%
  \BibitemOpen
  \bibfield  {author} {\bibinfo {author} {\bibfnamefont {L.}~\bibnamefont
  {Badesa}}, \bibinfo {author} {\bibfnamefont {G.}~\bibnamefont {Strbac}},
  \bibinfo {author} {\bibfnamefont {M.}~\bibnamefont {Magill}},\ and\ \bibinfo
  {author} {\bibfnamefont {B.}~\bibnamefont {Stojkovska}},\ }\bibfield  {title}
  {\bibinfo {title} {Ancillary services in {G}reat {B}ritain during the
  {COVID-19} lockdown: {A} glimpse of the carbon-free future},\ }\href
  {https://doi.org/10.1016/j.apenergy.2021.116500} {\bibfield  {journal}
  {\bibinfo  {journal} {Applied Energy}\ }\textbf {\bibinfo {volume} {285}},\
  \bibinfo {pages} {116500} (\bibinfo {year} {2021})}\BibitemShut {NoStop}%
\bibitem [{EPE(2021{\natexlab{a}})}]{EPEX}%
  \BibitemOpen
  \href@noop {} {\bibinfo {title} {{European Power Exchange (EPEX SPOT)}}}
  (\bibinfo {year} {2021}{\natexlab{a}}),\ \bibinfo {note} {{M}arket data
  \href{https://www.epexspot.com/en/market-data}{https://www.epexspot.com/en/market-data}}\BibitemShut
  {NoStop}%
\bibitem [{Nor(2021)}]{NordPool}%
  \BibitemOpen
  \href@noop {} {\bibinfo {title} {{Historical Market Data (Nord Pool)}}}
  (\bibinfo {year} {2021}),\ \bibinfo {note} {{M}arket data
  \href{https://www.nordpoolgroup.com/historical-market-data/}{https://www.nordpoolgroup.com/historical-market-data/}}\BibitemShut
  {NoStop}%
\bibitem [{\citenamefont {Edenhofer}\ \emph {et~al.}(2013)\citenamefont
  {Edenhofer}, \citenamefont {Hirth}, \citenamefont {Knopf}, \citenamefont
  {Pahle}, \citenamefont {Schlömer}, \citenamefont {Schmid},\ and\
  \citenamefont {Ueckerdt}}]{Edenhofer2013}%
  \BibitemOpen
  \bibfield  {author} {\bibinfo {author} {\bibfnamefont {O.}~\bibnamefont
  {Edenhofer}}, \bibinfo {author} {\bibfnamefont {L.}~\bibnamefont {Hirth}},
  \bibinfo {author} {\bibfnamefont {B.}~\bibnamefont {Knopf}}, \bibinfo
  {author} {\bibfnamefont {M.}~\bibnamefont {Pahle}}, \bibinfo {author}
  {\bibfnamefont {S.}~\bibnamefont {Schlömer}}, \bibinfo {author}
  {\bibfnamefont {E.}~\bibnamefont {Schmid}},\ and\ \bibinfo {author}
  {\bibfnamefont {F.}~\bibnamefont {Ueckerdt}},\ }\bibfield  {title} {\bibinfo
  {title} {On the economics of renewable energy sources},\ }\href
  {https://doi.org/10.1016/j.eneco.2013.09.015} {\bibfield  {journal} {\bibinfo
   {journal} {Energy Economics}\ }\textbf {\bibinfo {volume} {40}},\ \bibinfo
  {pages} {S12} (\bibinfo {year} {2013})}\BibitemShut {NoStop}%
\bibitem [{\citenamefont {Spodniak}\ \emph {et~al.}(2021)\citenamefont
  {Spodniak}, \citenamefont {Ollikka},\ and\ \citenamefont
  {Honkapuro}}]{Spodniak2021}%
  \BibitemOpen
  \bibfield  {author} {\bibinfo {author} {\bibfnamefont {P.}~\bibnamefont
  {Spodniak}}, \bibinfo {author} {\bibfnamefont {K.}~\bibnamefont {Ollikka}},\
  and\ \bibinfo {author} {\bibfnamefont {S.}~\bibnamefont {Honkapuro}},\
  }\bibfield  {title} {\bibinfo {title} {The impact of wind power and
  electricity demand on the relevance of different short-term electricity
  markets: {T}he {N}ordic case},\ }\href
  {https://doi.org/10.1016/j.apenergy.2020.116063} {\bibfield  {journal}
  {\bibinfo  {journal} {Applied Energy}\ }\textbf {\bibinfo {volume} {283}},\
  \bibinfo {pages} {116063} (\bibinfo {year} {2021})}\BibitemShut {NoStop}%
\bibitem [{\citenamefont {Braun}\ and\ \citenamefont
  {Brunner}(2018)}]{Braun2018}%
  \BibitemOpen
  \bibfield  {author} {\bibinfo {author} {\bibfnamefont {S.~M.}\ \bibnamefont
  {Braun}}\ and\ \bibinfo {author} {\bibfnamefont {C.}~\bibnamefont
  {Brunner}},\ }\bibfield  {title} {\bibinfo {title} {Price sensitivity of
  hourly day-ahead and quarter-hourly intraday auctions in germany},\ }\href
  {https://doi.org/10.1007/s12398-018-0228-0} {\bibfield  {journal} {\bibinfo
  {journal} {Zeitschrift f{\"u}r {E}nergiewirtschaft}\ }\textbf {\bibinfo
  {volume} {42}},\ \bibinfo {pages} {257} (\bibinfo {year} {2018})}\BibitemShut
  {NoStop}%
\bibitem [{\citenamefont {Secretariat}(2021)}]{REN2021}%
  \BibitemOpen
  \bibfield  {author} {\bibinfo {author} {\bibfnamefont {R.}~\bibnamefont
  {Secretariat}},\ }\href
  {https://www.ren21.net/wp-content/uploads/2019/05/GSR2021_Full_Report.pdf}
  {\emph {\bibinfo {title} {Renewables 2021 {G}lobal {S}tatus {R}eport}}}\
  (\bibinfo  {publisher} {REN21},\ \bibinfo {year} {2021})\ \bibinfo {note}
  {{ISBN}: 978-3-948393-03-8}\BibitemShut {NoStop}%
\bibitem [{\citenamefont {Rodr{\'\i}guez-Molina}\ \emph
  {et~al.}(2014)\citenamefont {Rodr{\'\i}guez-Molina}, \citenamefont
  {Mart{\'\i}nez-N{\'u}{\~n}ez}, \citenamefont {Mart{\'\i}nez},\ and\
  \citenamefont {P{\'e}rez-Aguiar}}]{Rodriguez2014}%
  \BibitemOpen
  \bibfield  {author} {\bibinfo {author} {\bibfnamefont {J.}~\bibnamefont
  {Rodr{\'\i}guez-Molina}}, \bibinfo {author} {\bibfnamefont {M.}~\bibnamefont
  {Mart{\'\i}nez-N{\'u}{\~n}ez}}, \bibinfo {author} {\bibfnamefont {J.-F.}\
  \bibnamefont {Mart{\'\i}nez}},\ and\ \bibinfo {author} {\bibfnamefont
  {W.}~\bibnamefont {P{\'e}rez-Aguiar}},\ }\bibfield  {title} {\bibinfo {title}
  {Business models in the smart grid: {C}hallenges, opportunities and proposals
  for prosumer profitability},\ }\href {https://doi.org/10.3390/en7096142}
  {\bibfield  {journal} {\bibinfo  {journal} {Energies}\ }\textbf {\bibinfo
  {volume} {7}},\ \bibinfo {pages} {6142} (\bibinfo {year} {2014})}\BibitemShut
  {NoStop}%
\bibitem [{\citenamefont {Conejo}\ and\ \citenamefont
  {Sioshansi}(2018)}]{Conejo2018}%
  \BibitemOpen
  \bibfield  {author} {\bibinfo {author} {\bibfnamefont {A.~J.}\ \bibnamefont
  {Conejo}}\ and\ \bibinfo {author} {\bibfnamefont {R.}~\bibnamefont
  {Sioshansi}},\ }\bibfield  {title} {\bibinfo {title} {Rethinking restructured
  electricity market design: {L}essons learned and future needs},\ }\href
  {https://doi.org/10.1016/j.ijepes.2017.12.014} {\bibfield  {journal}
  {\bibinfo  {journal} {International Journal of Electrical Power \& Energy
  Systems}\ }\textbf {\bibinfo {volume} {98}},\ \bibinfo {pages} {520}
  (\bibinfo {year} {2018})}\BibitemShut {NoStop}%
\bibitem [{\citenamefont {Macedo}\ \emph {et~al.}(2020)\citenamefont {Macedo},
  \citenamefont {Marques},\ and\ \citenamefont {Damette}}]{Macedo2020}%
  \BibitemOpen
  \bibfield  {author} {\bibinfo {author} {\bibfnamefont {D.~P.}\ \bibnamefont
  {Macedo}}, \bibinfo {author} {\bibfnamefont {A.~C.}\ \bibnamefont
  {Marques}},\ and\ \bibinfo {author} {\bibfnamefont {O.}~\bibnamefont
  {Damette}},\ }\bibfield  {title} {\bibinfo {title} {The impact of the
  integration of renewable energy sources in the electricity price formation:
  is the {M}erit-{O}rder {E}ffect occurring in {P}ortugal?},\ }\href
  {https://doi.org/10.1016/j.jup.2020.101080} {\bibfield  {journal} {\bibinfo
  {journal} {Utilities Policy}\ }\textbf {\bibinfo {volume} {66}},\ \bibinfo
  {pages} {101080} (\bibinfo {year} {2020})}\BibitemShut {NoStop}%
\bibitem [{\citenamefont {Peters}\ \emph {et~al.}(2020)\citenamefont {Peters},
  \citenamefont {Völker}, \citenamefont {Schuldt},\ and\ \citenamefont {von
  Maydell}}]{Peters2020}%
  \BibitemOpen
  \bibfield  {author} {\bibinfo {author} {\bibfnamefont {D.}~\bibnamefont
  {Peters}}, \bibinfo {author} {\bibfnamefont {R.}~\bibnamefont {Völker}},
  \bibinfo {author} {\bibfnamefont {F.}~\bibnamefont {Schuldt}},\ and\ \bibinfo
  {author} {\bibfnamefont {K.}~\bibnamefont {von Maydell}},\ }\bibfield
  {title} {\bibinfo {title} {Are standard load profiles suitable for modern
  electricity grid models?},\ }in\ \href
  {https://doi.org/10.1109/EEM49802.2020.9221967} {\emph {\bibinfo {booktitle}
  {2020 17th International Conference on the European Energy Market (EEM)}}}\
  (\bibinfo {year} {2020})\ pp.\ \bibinfo {pages} {1--6}\BibitemShut {NoStop}%
\bibitem [{\citenamefont {Sensfu{\ss}}\ \emph {et~al.}(2008)\citenamefont
  {Sensfu{\ss}}, \citenamefont {Ragwitz},\ and\ \citenamefont
  {Genoese}}]{Sensfuss2008}%
  \BibitemOpen
  \bibfield  {author} {\bibinfo {author} {\bibfnamefont {F.}~\bibnamefont
  {Sensfu{\ss}}}, \bibinfo {author} {\bibfnamefont {M.}~\bibnamefont
  {Ragwitz}},\ and\ \bibinfo {author} {\bibfnamefont {M.}~\bibnamefont
  {Genoese}},\ }\bibfield  {title} {\bibinfo {title} {The merit-order effect:
  {A} detailed analysis of the price effect of renewable electricity generation
  on spot market prices in {G}ermany},\ }\href
  {https://doi.org/10.1016/j.enpol.2008.03.035} {\bibfield  {journal} {\bibinfo
   {journal} {Energy Policy}\ }\textbf {\bibinfo {volume} {36}},\ \bibinfo
  {pages} {3086} (\bibinfo {year} {2008})}\BibitemShut {NoStop}%
\bibitem [{\citenamefont {Infield}\ and\ \citenamefont
  {Freris}(2020)}]{Infield2020}%
  \BibitemOpen
  \bibfield  {author} {\bibinfo {author} {\bibfnamefont {D.}~\bibnamefont
  {Infield}}\ and\ \bibinfo {author} {\bibfnamefont {L.}~\bibnamefont
  {Freris}},\ }\href@noop {} {\emph {\bibinfo {title} {Renewable Energy in
  Power Systems}}},\ \bibinfo {edition} {2nd}\ ed.\ (\bibinfo  {publisher}
  {John Wiley \& Sons, Wiltshire, UK},\ \bibinfo {year} {2020})\BibitemShut
  {NoStop}%
\bibitem [{\citenamefont {Cludius}\ \emph {et~al.}(2014)\citenamefont
  {Cludius}, \citenamefont {Hermann}, \citenamefont {Matthes},\ and\
  \citenamefont {Graichen}}]{Cludius2014}%
  \BibitemOpen
  \bibfield  {author} {\bibinfo {author} {\bibfnamefont {J.}~\bibnamefont
  {Cludius}}, \bibinfo {author} {\bibfnamefont {H.}~\bibnamefont {Hermann}},
  \bibinfo {author} {\bibfnamefont {F.~C.}\ \bibnamefont {Matthes}},\ and\
  \bibinfo {author} {\bibfnamefont {V.}~\bibnamefont {Graichen}},\ }\bibfield
  {title} {\bibinfo {title} {The merit order effect of wind and photovoltaic
  electricity generation in {G}ermany 2008--2016: {E}stimation and
  distributional implications},\ }\href
  {https://doi.org/10.1016/j.eneco.2014.04.020} {\bibfield  {journal} {\bibinfo
   {journal} {Energy Economics}\ }\textbf {\bibinfo {volume} {44}},\ \bibinfo
  {pages} {302} (\bibinfo {year} {2014})}\BibitemShut {NoStop}%
\bibitem [{\citenamefont {Kiesel}\ and\ \citenamefont
  {Paraschiv}(2017)}]{Kiesel2017}%
  \BibitemOpen
  \bibfield  {author} {\bibinfo {author} {\bibfnamefont {R.}~\bibnamefont
  {Kiesel}}\ and\ \bibinfo {author} {\bibfnamefont {F.}~\bibnamefont
  {Paraschiv}},\ }\bibfield  {title} {\bibinfo {title} {Econometric analysis of
  15-minute intraday electricity prices},\ }\href
  {https://doi.org/10.1016/j.eneco.2017.03.002} {\bibfield  {journal} {\bibinfo
   {journal} {Energy Economics}\ }\textbf {\bibinfo {volume} {64}},\ \bibinfo
  {pages} {77} (\bibinfo {year} {2017})}\BibitemShut {NoStop}%
\bibitem [{\citenamefont {Hirth}(2013)}]{Hirth2013}%
  \BibitemOpen
  \bibfield  {author} {\bibinfo {author} {\bibfnamefont {L.}~\bibnamefont
  {Hirth}},\ }\bibfield  {title} {\bibinfo {title} {The market value of
  variable renewables: {T}he effect of solar wind power variability on their
  relative price},\ }\href {https://doi.org/10.1016/j.eneco.2013.02.004}
  {\bibfield  {journal} {\bibinfo  {journal} {Energy Economics}\ }\textbf
  {\bibinfo {volume} {38}},\ \bibinfo {pages} {218} (\bibinfo {year}
  {2013})}\BibitemShut {NoStop}%
\bibitem [{\citenamefont {Ketterer}(2014)}]{Ketterer2014}%
  \BibitemOpen
  \bibfield  {author} {\bibinfo {author} {\bibfnamefont {J.~C.}\ \bibnamefont
  {Ketterer}},\ }\bibfield  {title} {\bibinfo {title} {The impact of wind power
  generation on the electricity price in {G}ermany},\ }\href
  {https://doi.org/10.1016/j.eneco.2014.04.003} {\bibfield  {journal} {\bibinfo
   {journal} {Energy Economics}\ }\textbf {\bibinfo {volume} {44}},\ \bibinfo
  {pages} {270} (\bibinfo {year} {2014})}\BibitemShut {NoStop}%
\bibitem [{\citenamefont {Hirth}\ and\ \citenamefont
  {Ziegenhagen}(2015)}]{Hirth2015}%
  \BibitemOpen
  \bibfield  {author} {\bibinfo {author} {\bibfnamefont {L.}~\bibnamefont
  {Hirth}}\ and\ \bibinfo {author} {\bibfnamefont {I.}~\bibnamefont
  {Ziegenhagen}},\ }\bibfield  {title} {\bibinfo {title} {Balancing power and
  variable renewables: {T}hree links},\ }\href
  {https://doi.org/10.1016/j.rser.2015.04.180} {\bibfield  {journal} {\bibinfo
  {journal} {Renewable and Sustainable Energy Reviews}\ }\textbf {\bibinfo
  {volume} {50}},\ \bibinfo {pages} {1035} (\bibinfo {year}
  {2015})}\BibitemShut {NoStop}%
\bibitem [{\citenamefont {Kremer}\ \emph {et~al.}(2020)\citenamefont {Kremer},
  \citenamefont {Kiesel},\ and\ \citenamefont {Paraschiv}}]{Kremer2020}%
  \BibitemOpen
  \bibfield  {author} {\bibinfo {author} {\bibfnamefont {M.}~\bibnamefont
  {Kremer}}, \bibinfo {author} {\bibfnamefont {R.}~\bibnamefont {Kiesel}},\
  and\ \bibinfo {author} {\bibfnamefont {F.}~\bibnamefont {Paraschiv}},\
  }\bibfield  {title} {\bibinfo {title} {Intraday electricity pricing of night
  contracts},\ }\bibfield  {journal} {\bibinfo  {journal} {Energies}\ }\textbf
  {\bibinfo {volume} {13}},\ \href {https://doi.org/10.3390/en13174501}
  {10.3390/en13174501} (\bibinfo {year} {2020})\BibitemShut {NoStop}%
\bibitem [{\citenamefont {{Crespo Cuaresma}}\ \emph {et~al.}(2004)\citenamefont
  {{Crespo Cuaresma}}, \citenamefont {Hlouskova}, \citenamefont {Kossmeier},\
  and\ \citenamefont {Obersteiner}}]{Cuaresma2004}%
  \BibitemOpen
  \bibfield  {author} {\bibinfo {author} {\bibfnamefont {J.}~\bibnamefont
  {{Crespo Cuaresma}}}, \bibinfo {author} {\bibfnamefont {J.}~\bibnamefont
  {Hlouskova}}, \bibinfo {author} {\bibfnamefont {S.}~\bibnamefont
  {Kossmeier}},\ and\ \bibinfo {author} {\bibfnamefont {M.}~\bibnamefont
  {Obersteiner}},\ }\bibfield  {title} {\bibinfo {title} {Forecasting
  electricity spot-prices using linear univariate time-series models},\ }\href
  {https://doi.org/10.1016/S0306-2619(03)00096-5} {\bibfield  {journal}
  {\bibinfo  {journal} {Applied Energy}\ }\textbf {\bibinfo {volume} {77}},\
  \bibinfo {pages} {87} (\bibinfo {year} {2004})}\BibitemShut {NoStop}%
\bibitem [{\citenamefont {Schwartz}(1997)}]{Schwartz1997}%
  \BibitemOpen
  \bibfield  {author} {\bibinfo {author} {\bibfnamefont {E.~S.}\ \bibnamefont
  {Schwartz}},\ }\bibfield  {title} {\bibinfo {title} {The stochastic behavior
  of commodity prices: Implications for valuation and hedging},\ }\href
  {https://doi.org/10.1111/j.1540-6261.1997.tb02721.x} {\bibfield  {journal}
  {\bibinfo  {journal} {The Journal of Finance}\ }\textbf {\bibinfo {volume}
  {52}},\ \bibinfo {pages} {923} (\bibinfo {year} {1997})}\BibitemShut
  {NoStop}%
\bibitem [{\citenamefont {Weron}\ \emph
  {et~al.}(2004{\natexlab{a}})\citenamefont {Weron}, \citenamefont {Simonsen},\
  and\ \citenamefont {Wilman}}]{Weron2004b}%
  \BibitemOpen
  \bibfield  {author} {\bibinfo {author} {\bibfnamefont {R.}~\bibnamefont
  {Weron}}, \bibinfo {author} {\bibfnamefont {I.}~\bibnamefont {Simonsen}},\
  and\ \bibinfo {author} {\bibfnamefont {P.}~\bibnamefont {Wilman}},\
  }\bibfield  {title} {\bibinfo {title} {Modeling highly volatile and seasonal
  markets: {E}vidence from the {N}ord {P}ool electricity market},\ }in\ \href
  {https://doi.org/10.1007/978-4-431-53947-6_25} {\emph {\bibinfo {booktitle}
  {The Application of Econophysics}}},\ \bibinfo {editor} {edited by\ \bibinfo
  {editor} {\bibfnamefont {H.}~\bibnamefont {Takayasu}}}\ (\bibinfo
  {publisher} {Springer Japan},\ \bibinfo {address} {Tokyo},\ \bibinfo {year}
  {2004})\ pp.\ \bibinfo {pages} {182--191},\ \bibinfo {note} {{ISBN}:
  978-4-431-53947-6}\BibitemShut {NoStop}%
\bibitem [{\citenamefont {Han}\ \emph {et~al.}(2021)\citenamefont {Han},
  \citenamefont {Hilger}, \citenamefont {Mix}, \citenamefont {B{\"o}ttcher},
  \citenamefont {Reyers}, \citenamefont {Beck}, \citenamefont {Witthaut},\ and\
  \citenamefont {Rydin Gorj\~ao}}]{Han2021}%
  \BibitemOpen
  \bibfield  {author} {\bibinfo {author} {\bibfnamefont {C.}~\bibnamefont
  {Han}}, \bibinfo {author} {\bibfnamefont {H.}~\bibnamefont {Hilger}},
  \bibinfo {author} {\bibfnamefont {E.}~\bibnamefont {Mix}}, \bibinfo {author}
  {\bibfnamefont {{\relax Ph}.~C.}\ \bibnamefont {B{\"o}ttcher}}, \bibinfo
  {author} {\bibfnamefont {M.}~\bibnamefont {Reyers}}, \bibinfo {author}
  {\bibfnamefont {C.}~\bibnamefont {Beck}}, \bibinfo {author} {\bibfnamefont
  {D.}~\bibnamefont {Witthaut}},\ and\ \bibinfo {author} {\bibfnamefont
  {L.}~\bibnamefont {Rydin Gorj\~ao}},\ }\bibfield  {title} {\bibinfo {title}
  {Complexity and persistence of price time series of the european electricity
  spot market},\ }\href@noop {} {\bibfield  {journal} {\bibinfo  {journal}
  {\href{https://arxiv.org/abs/2112.03031}{preprint arXiv:2112.03031}}\ }
  (\bibinfo {year} {2021})}\BibitemShut {NoStop}%
\bibitem [{\citenamefont {Weron}(2002)}]{Weron2002}%
  \BibitemOpen
  \bibfield  {author} {\bibinfo {author} {\bibfnamefont {R.}~\bibnamefont
  {Weron}},\ }\bibfield  {title} {\bibinfo {title} {Measuring long-range
  dependence in electricity prices},\ }in\ \href
  {https://doi.org/10.1007/978-4-431-66993-7_12} {\emph {\bibinfo {booktitle}
  {Empirical Science of Financial Fluctuations}}},\ \bibinfo {editor} {edited
  by\ \bibinfo {editor} {\bibfnamefont {H.}~\bibnamefont {Takayasu}}}\
  (\bibinfo  {publisher} {Springer Japan},\ \bibinfo {address} {Tokyo},\
  \bibinfo {year} {2002})\ pp.\ \bibinfo {pages} {110--119},\ \bibinfo {note}
  {{ISBN}: 978-4-431-66993-7}\BibitemShut {NoStop}%
\bibitem [{\citenamefont {Kremer}\ \emph {et~al.}(2021)\citenamefont {Kremer},
  \citenamefont {Kiesel},\ and\ \citenamefont {Paraschiv}}]{Kremer2021}%
  \BibitemOpen
  \bibfield  {author} {\bibinfo {author} {\bibfnamefont {M.}~\bibnamefont
  {Kremer}}, \bibinfo {author} {\bibfnamefont {R.}~\bibnamefont {Kiesel}},\
  and\ \bibinfo {author} {\bibfnamefont {F.}~\bibnamefont {Paraschiv}},\
  }\bibfield  {title} {\bibinfo {title} {An econometric model for intraday
  electricity trading},\ }\href {https://doi.org/10.1098/rsta.2019.0624}
  {\bibfield  {journal} {\bibinfo  {journal} {Philosophical Transactions of the
  Royal Society A: Mathematical, Physical and Engineering Sciences}\ }\textbf
  {\bibinfo {volume} {379}},\ \bibinfo {pages} {20190624} (\bibinfo {year}
  {2021})}\BibitemShut {NoStop}%
\bibitem [{\citenamefont {Halbrügge}\ \emph {et~al.}(2021)\citenamefont
  {Halbrügge}, \citenamefont {Schott}, \citenamefont {Weibelzahl},
  \citenamefont {Buhl}, \citenamefont {Fridgen},\ and\ \citenamefont
  {Schöpf}}]{Halbruegge2021}%
  \BibitemOpen
  \bibfield  {author} {\bibinfo {author} {\bibfnamefont {S.}~\bibnamefont
  {Halbrügge}}, \bibinfo {author} {\bibfnamefont {P.}~\bibnamefont {Schott}},
  \bibinfo {author} {\bibfnamefont {M.}~\bibnamefont {Weibelzahl}}, \bibinfo
  {author} {\bibfnamefont {H.~U.}\ \bibnamefont {Buhl}}, \bibinfo {author}
  {\bibfnamefont {G.}~\bibnamefont {Fridgen}},\ and\ \bibinfo {author}
  {\bibfnamefont {M.}~\bibnamefont {Schöpf}},\ }\bibfield  {title} {\bibinfo
  {title} {How did the {G}erman and other {E}uropean electricity systems react
  to the {COVID-19} pandemic?},\ }\href
  {https://doi.org/10.1016/j.apenergy.2020.116370} {\bibfield  {journal}
  {\bibinfo  {journal} {Applied Energy}\ }\textbf {\bibinfo {volume} {285}},\
  \bibinfo {pages} {116370} (\bibinfo {year} {2021})}\BibitemShut {NoStop}%
\bibitem [{\citenamefont {Mandelbrot}\ and\ \citenamefont
  {Hudson}(2004)}]{Mandelbrot2004}%
  \BibitemOpen
  \bibfield  {author} {\bibinfo {author} {\bibfnamefont {B.~B.}\ \bibnamefont
  {Mandelbrot}}\ and\ \bibinfo {author} {\bibfnamefont {R.~L.}\ \bibnamefont
  {Hudson}},\ }\href@noop {} {\emph {\bibinfo {title} {{The (Mis)Behaviour of
  Markets}}}},\ \bibinfo {edition} {2nd}\ ed.\ (\bibinfo  {publisher} {Basic
  Books},\ \bibinfo {year} {2004})\ \bibinfo {note} {{ISBN}:
  978-0-465-04355-2}\BibitemShut {NoStop}%
\bibitem [{\citenamefont {Keles}\ \emph {et~al.}(2012)\citenamefont {Keles},
  \citenamefont {Genoese}, \citenamefont {Möst},\ and\ \citenamefont
  {Fichtner}}]{Keles2012}%
  \BibitemOpen
  \bibfield  {author} {\bibinfo {author} {\bibfnamefont {D.}~\bibnamefont
  {Keles}}, \bibinfo {author} {\bibfnamefont {M.}~\bibnamefont {Genoese}},
  \bibinfo {author} {\bibfnamefont {D.}~\bibnamefont {Möst}},\ and\ \bibinfo
  {author} {\bibfnamefont {W.}~\bibnamefont {Fichtner}},\ }\bibfield  {title}
  {\bibinfo {title} {Comparison of extended mean-reversion and time series
  models for electricity spot price simulation considering negative prices},\
  }\href {https://doi.org/10.1016/j.eneco.2011.08.012} {\bibfield  {journal}
  {\bibinfo  {journal} {Energy Economics}\ }\textbf {\bibinfo {volume} {34}},\
  \bibinfo {pages} {1012} (\bibinfo {year} {2012})}\BibitemShut {NoStop}%
\bibitem [{\citenamefont {Weron}(2014)}]{Weron2014}%
  \BibitemOpen
  \bibfield  {author} {\bibinfo {author} {\bibfnamefont {R.}~\bibnamefont
  {Weron}},\ }\bibfield  {title} {\bibinfo {title} {Electricity price
  forecasting: {A} review of the state-of-the-art with a look into the
  future},\ }\href {https://doi.org/10.1016/j.ijforecast.2014.08.008}
  {\bibfield  {journal} {\bibinfo  {journal} {International Journal of
  Forecasting}\ }\textbf {\bibinfo {volume} {30}},\ \bibinfo {pages} {1030}
  (\bibinfo {year} {2014})}\BibitemShut {NoStop}%
\bibitem [{\citenamefont {Weron}\ \emph
  {et~al.}(2004{\natexlab{b}})\citenamefont {Weron}, \citenamefont
  {Bierbrauer},\ and\ \citenamefont {Trück}}]{Weron2004}%
  \BibitemOpen
  \bibfield  {author} {\bibinfo {author} {\bibfnamefont {R.}~\bibnamefont
  {Weron}}, \bibinfo {author} {\bibfnamefont {M.}~\bibnamefont {Bierbrauer}},\
  and\ \bibinfo {author} {\bibfnamefont {S.}~\bibnamefont {Trück}},\
  }\bibfield  {title} {\bibinfo {title} {Modeling electricity prices: jump
  diffusion and regime switching},\ }\href
  {https://doi.org/10.1016/j.physa.2004.01.008} {\bibfield  {journal} {\bibinfo
   {journal} {Physica A: Statistical Mechanics and its Applications}\ }\textbf
  {\bibinfo {volume} {336}},\ \bibinfo {pages} {39} (\bibinfo {year}
  {2004}{\natexlab{b}})},\ \bibinfo {note} {proceedings of the XVIII Max Born
  Symposium ``Statistical Physics outside Physics''}\BibitemShut {NoStop}%
\bibitem [{\citenamefont {Huisman}\ and\ \citenamefont
  {Mahieu}(2003)}]{Huisman2003}%
  \BibitemOpen
  \bibfield  {author} {\bibinfo {author} {\bibfnamefont {R.}~\bibnamefont
  {Huisman}}\ and\ \bibinfo {author} {\bibfnamefont {R.}~\bibnamefont
  {Mahieu}},\ }\bibfield  {title} {\bibinfo {title} {Regime jumps in
  electricity prices},\ }\href {https://doi.org/10.1016/S0140-9883(03)00041-0}
  {\bibfield  {journal} {\bibinfo  {journal} {Energy Economics}\ }\textbf
  {\bibinfo {volume} {25}},\ \bibinfo {pages} {425} (\bibinfo {year}
  {2003})}\BibitemShut {NoStop}%
\bibitem [{\citenamefont {Janczura}\ \emph {et~al.}(2013)\citenamefont
  {Janczura}, \citenamefont {Trück}, \citenamefont {Weron},\ and\
  \citenamefont {Wolff}}]{Janczura2013}%
  \BibitemOpen
  \bibfield  {author} {\bibinfo {author} {\bibfnamefont {J.}~\bibnamefont
  {Janczura}}, \bibinfo {author} {\bibfnamefont {S.}~\bibnamefont {Trück}},
  \bibinfo {author} {\bibfnamefont {R.}~\bibnamefont {Weron}},\ and\ \bibinfo
  {author} {\bibfnamefont {R.~C.}\ \bibnamefont {Wolff}},\ }\bibfield  {title}
  {\bibinfo {title} {Identifying spikes and seasonal components in electricity
  spot price data: {A} guide to robust modeling},\ }\href
  {https://doi.org/10.1016/j.eneco.2013.03.013} {\bibfield  {journal} {\bibinfo
   {journal} {Energy Economics}\ }\textbf {\bibinfo {volume} {38}},\ \bibinfo
  {pages} {96} (\bibinfo {year} {2013})}\BibitemShut {NoStop}%
\bibitem [{\citenamefont {Kiesel}\ \emph {et~al.}(2019)\citenamefont {Kiesel},
  \citenamefont {Paraschiv},\ and\ \citenamefont
  {S{\ae}ther{\o}}}]{Kiesel2019}%
  \BibitemOpen
  \bibfield  {author} {\bibinfo {author} {\bibfnamefont {R.}~\bibnamefont
  {Kiesel}}, \bibinfo {author} {\bibfnamefont {F.}~\bibnamefont {Paraschiv}},\
  and\ \bibinfo {author} {\bibfnamefont {A.}~\bibnamefont {S{\ae}ther{\o}}},\
  }\bibfield  {title} {\bibinfo {title} {On the construction of hourly price
  forward curves for electricity prices},\ }\href
  {https://doi.org/10.1007/s10287-018-0300-6} {\bibfield  {journal} {\bibinfo
  {journal} {Computational Management Science}\ }\textbf {\bibinfo {volume}
  {16}},\ \bibinfo {pages} {345} (\bibinfo {year} {2019})}\BibitemShut
  {NoStop}%
\bibitem [{\citenamefont {Norouzzadeh}\ \emph {et~al.}(2007)\citenamefont
  {Norouzzadeh}, \citenamefont {Dullaert},\ and\ \citenamefont
  {Rahmani}}]{Norouzzadeh2007}%
  \BibitemOpen
  \bibfield  {author} {\bibinfo {author} {\bibfnamefont {P.}~\bibnamefont
  {Norouzzadeh}}, \bibinfo {author} {\bibfnamefont {W.}~\bibnamefont
  {Dullaert}},\ and\ \bibinfo {author} {\bibfnamefont {B.}~\bibnamefont
  {Rahmani}},\ }\bibfield  {title} {\bibinfo {title} {Anti-correlation and
  multifractal features of {S}pain electricity spot market},\ }\href
  {https://doi.org/10.1016/j.physa.2007.02.087} {\bibfield  {journal} {\bibinfo
   {journal} {Physica A: Statistical Mechanics and its Applications}\ }\textbf
  {\bibinfo {volume} {380}},\ \bibinfo {pages} {333} (\bibinfo {year}
  {2007})}\BibitemShut {NoStop}%
\bibitem [{\citenamefont {Alvarez-Ramirez}\ and\ \citenamefont
  {Escarela-Perez}(2010)}]{AlvarezRamirez2010}%
  \BibitemOpen
  \bibfield  {author} {\bibinfo {author} {\bibfnamefont {J.}~\bibnamefont
  {Alvarez-Ramirez}}\ and\ \bibinfo {author} {\bibfnamefont {R.}~\bibnamefont
  {Escarela-Perez}},\ }\bibfield  {title} {\bibinfo {title} {Time-dependent
  correlations in electricity markets},\ }\href
  {https://doi.org/10.1016/j.eneco.2009.05.008} {\bibfield  {journal} {\bibinfo
   {journal} {Energy Economics}\ }\textbf {\bibinfo {volume} {32}},\ \bibinfo
  {pages} {269} (\bibinfo {year} {2010})}\BibitemShut {NoStop}%
\bibitem [{\citenamefont {Wang}\ \emph
  {et~al.}(2013{\natexlab{a}})\citenamefont {Wang}, \citenamefont {ping Liao},
  \citenamefont {hui Li}, \citenamefont {chun Li},\ and\ \citenamefont {jun
  Zhou}}]{Wang2013a}%
  \BibitemOpen
  \bibfield  {author} {\bibinfo {author} {\bibfnamefont {F.}~\bibnamefont
  {Wang}}, \bibinfo {author} {\bibfnamefont {G.}~\bibnamefont {ping Liao}},
  \bibinfo {author} {\bibfnamefont {J.}~\bibnamefont {hui Li}}, \bibinfo
  {author} {\bibfnamefont {X.}~\bibnamefont {chun Li}},\ and\ \bibinfo {author}
  {\bibfnamefont {T.}~\bibnamefont {jun Zhou}},\ }\bibfield  {title} {\bibinfo
  {title} {Multifractal detrended fluctuation analysis for clustering
  structures of electricity price periods},\ }\href
  {https://doi.org/10.1016/j.physa.2013.07.039} {\bibfield  {journal} {\bibinfo
   {journal} {Physica A: Statistical Mechanics and its Applications}\ }\textbf
  {\bibinfo {volume} {392}},\ \bibinfo {pages} {5723} (\bibinfo {year}
  {2013}{\natexlab{a}})}\BibitemShut {NoStop}%
\bibitem [{\citenamefont {Wang}\ \emph
  {et~al.}(2013{\natexlab{b}})\citenamefont {Wang}, \citenamefont {Liao},
  \citenamefont {Zhou},\ and\ \citenamefont {Shi}}]{Wang2013b}%
  \BibitemOpen
  \bibfield  {author} {\bibinfo {author} {\bibfnamefont {F.}~\bibnamefont
  {Wang}}, \bibinfo {author} {\bibfnamefont {G.-p.}\ \bibnamefont {Liao}},
  \bibinfo {author} {\bibfnamefont {X.-y.}\ \bibnamefont {Zhou}},\ and\
  \bibinfo {author} {\bibfnamefont {W.}~\bibnamefont {Shi}},\ }\bibfield
  {title} {\bibinfo {title} {Multifractal detrended cross-correlation analysis
  for power markets},\ }\href {https://doi.org/10.1007/s11071-012-0718-2}
  {\bibfield  {journal} {\bibinfo  {journal} {Nonlinear Dynamics}\ }\textbf
  {\bibinfo {volume} {72}},\ \bibinfo {pages} {353} (\bibinfo {year}
  {2013}{\natexlab{b}})}\BibitemShut {NoStop}%
\bibitem [{\citenamefont {Ergemen}\ \emph {et~al.}(2016)\citenamefont
  {Ergemen}, \citenamefont {Haldrup},\ and\ \citenamefont
  {Rodríguez-Caballero}}]{Ergemen2016}%
  \BibitemOpen
  \bibfield  {author} {\bibinfo {author} {\bibfnamefont {Y.~E.}\ \bibnamefont
  {Ergemen}}, \bibinfo {author} {\bibfnamefont {N.}~\bibnamefont {Haldrup}},\
  and\ \bibinfo {author} {\bibfnamefont {C.~V.}\ \bibnamefont
  {Rodríguez-Caballero}},\ }\bibfield  {title} {\bibinfo {title} {Common
  long-range dependence in a panel of hourly {N}ord {P}ool electricity prices
  and loads},\ }\href {https://doi.org/10.1016/j.eneco.2016.09.008} {\bibfield
   {journal} {\bibinfo  {journal} {Energy Economics}\ }\textbf {\bibinfo
  {volume} {60}},\ \bibinfo {pages} {79} (\bibinfo {year} {2016})}\BibitemShut
  {NoStop}%
\bibitem [{\citenamefont {Tan}\ \emph {et~al.}(2010)\citenamefont {Tan},
  \citenamefont {Zhang}, \citenamefont {Wang},\ and\ \citenamefont
  {Xu}}]{Tan2010}%
  \BibitemOpen
  \bibfield  {author} {\bibinfo {author} {\bibfnamefont {Z.}~\bibnamefont
  {Tan}}, \bibinfo {author} {\bibfnamefont {J.}~\bibnamefont {Zhang}}, \bibinfo
  {author} {\bibfnamefont {J.}~\bibnamefont {Wang}},\ and\ \bibinfo {author}
  {\bibfnamefont {J.}~\bibnamefont {Xu}},\ }\bibfield  {title} {\bibinfo
  {title} {Day-ahead electricity price forecasting using wavelet transform
  combined with {ARIMA} and {GARCH} models},\ }\href
  {https://doi.org/10.1016/j.apenergy.2010.05.012} {\bibfield  {journal}
  {\bibinfo  {journal} {Applied Energy}\ }\textbf {\bibinfo {volume} {87}},\
  \bibinfo {pages} {3606} (\bibinfo {year} {2010})}\BibitemShut {NoStop}%
\bibitem [{\citenamefont {Liu}\ and\ \citenamefont {Shi}(2013)}]{Liu2013}%
  \BibitemOpen
  \bibfield  {author} {\bibinfo {author} {\bibfnamefont {H.}~\bibnamefont
  {Liu}}\ and\ \bibinfo {author} {\bibfnamefont {J.}~\bibnamefont {Shi}},\
  }\bibfield  {title} {\bibinfo {title} {Applying {ARMA}--{GARCH} approaches to
  forecasting short-term electricity prices},\ }\href
  {https://doi.org/10.1016/j.eneco.2013.02.006} {\bibfield  {journal} {\bibinfo
   {journal} {Energy Economics}\ }\textbf {\bibinfo {volume} {37}},\ \bibinfo
  {pages} {152} (\bibinfo {year} {2013})}\BibitemShut {NoStop}%
\bibitem [{\citenamefont {Deng}\ \emph {et~al.}(2020)\citenamefont {Deng},
  \citenamefont {Song},\ and\ \citenamefont {Zio}}]{Deng2020}%
  \BibitemOpen
  \bibfield  {author} {\bibinfo {author} {\bibfnamefont {J.}~\bibnamefont
  {Deng}}, \bibinfo {author} {\bibfnamefont {W.}~\bibnamefont {Song}},\ and\
  \bibinfo {author} {\bibfnamefont {E.}~\bibnamefont {Zio}},\ }\bibfield
  {title} {\bibinfo {title} {A discrete increment model for electricity price
  forecasting based on fractional brownian motion},\ }\href
  {https://doi.org/10.1109/ACCESS.2020.3008797} {\bibfield  {journal} {\bibinfo
   {journal} {IEEE Access}\ }\textbf {\bibinfo {volume} {8}},\ \bibinfo {pages}
  {130762} (\bibinfo {year} {2020})}\BibitemShut {NoStop}%
\bibitem [{\citenamefont {Grassberger}\ and\ \citenamefont
  {Procaccia}(1984)}]{Grassberger1984}%
  \BibitemOpen
  \bibfield  {author} {\bibinfo {author} {\bibfnamefont {P.}~\bibnamefont
  {Grassberger}}\ and\ \bibinfo {author} {\bibfnamefont {I.}~\bibnamefont
  {Procaccia}},\ }\bibfield  {title} {\bibinfo {title} {Dimensions and
  entropies of strange attractors from a fluctuating dynamics approach},\
  }\href {https://doi.org/10.1016/0167-2789(84)90269-0} {\bibfield  {journal}
  {\bibinfo  {journal} {Physica D: Nonlinear Phenomena}\ }\textbf {\bibinfo
  {volume} {13}},\ \bibinfo {pages} {34} (\bibinfo {year} {1984})}\BibitemShut
  {NoStop}%
\bibitem [{\citenamefont {Halsey}\ \emph {et~al.}(1986)\citenamefont {Halsey},
  \citenamefont {Jensen}, \citenamefont {Kadanoff}, \citenamefont {Procaccia},\
  and\ \citenamefont {Shraiman}}]{Halsey1986}%
  \BibitemOpen
  \bibfield  {author} {\bibinfo {author} {\bibfnamefont {T.~C.}\ \bibnamefont
  {Halsey}}, \bibinfo {author} {\bibfnamefont {M.~H.}\ \bibnamefont {Jensen}},
  \bibinfo {author} {\bibfnamefont {L.~P.}\ \bibnamefont {Kadanoff}}, \bibinfo
  {author} {\bibfnamefont {I.}~\bibnamefont {Procaccia}},\ and\ \bibinfo
  {author} {\bibfnamefont {B.~I.}\ \bibnamefont {Shraiman}},\ }\bibfield
  {title} {\bibinfo {title} {Fractal measures and their singularities: The
  characterization of strange sets},\ }\href
  {https://doi.org/10.1103/PhysRevA.33.1141} {\bibfield  {journal} {\bibinfo
  {journal} {Physical Review A}\ }\textbf {\bibinfo {volume} {33}},\ \bibinfo
  {pages} {1141} (\bibinfo {year} {1986})}\BibitemShut {NoStop}%
\bibitem [{\citenamefont {Peng}\ \emph {et~al.}(1994)\citenamefont {Peng},
  \citenamefont {Buldyrev}, \citenamefont {Havlin}, \citenamefont {Simons},
  \citenamefont {Stanley},\ and\ \citenamefont {Goldberger}}]{Peng1994}%
  \BibitemOpen
  \bibfield  {author} {\bibinfo {author} {\bibfnamefont {C.-K.}\ \bibnamefont
  {Peng}}, \bibinfo {author} {\bibfnamefont {S.~V.}\ \bibnamefont {Buldyrev}},
  \bibinfo {author} {\bibfnamefont {S.}~\bibnamefont {Havlin}}, \bibinfo
  {author} {\bibfnamefont {M.}~\bibnamefont {Simons}}, \bibinfo {author}
  {\bibfnamefont {H.~E.}\ \bibnamefont {Stanley}},\ and\ \bibinfo {author}
  {\bibfnamefont {A.~L.}\ \bibnamefont {Goldberger}},\ }\bibfield  {title}
  {\bibinfo {title} {Mosaic organization of {DNA} nucleotides},\ }\href
  {https://doi.org/10.1103/PhysRevE.49.1685} {\bibfield  {journal} {\bibinfo
  {journal} {Physical Review E}\ }\textbf {\bibinfo {volume} {49}},\ \bibinfo
  {pages} {1685} (\bibinfo {year} {1994})}\BibitemShut {NoStop}%
\bibitem [{\citenamefont {Kantelhardt}\ \emph {et~al.}(2002)\citenamefont
  {Kantelhardt}, \citenamefont {Zschiegner}, \citenamefont {Koscielny-Bunde},
  \citenamefont {Havlin}, \citenamefont {Bunde},\ and\ \citenamefont
  {Stanley}}]{Kantelhardt2002}%
  \BibitemOpen
  \bibfield  {author} {\bibinfo {author} {\bibfnamefont {J.~W.}\ \bibnamefont
  {Kantelhardt}}, \bibinfo {author} {\bibfnamefont {S.~A.}\ \bibnamefont
  {Zschiegner}}, \bibinfo {author} {\bibfnamefont {E.}~\bibnamefont
  {Koscielny-Bunde}}, \bibinfo {author} {\bibfnamefont {S.}~\bibnamefont
  {Havlin}}, \bibinfo {author} {\bibfnamefont {A.}~\bibnamefont {Bunde}},\ and\
  \bibinfo {author} {\bibfnamefont {H.}~\bibnamefont {Stanley}},\ }\bibfield
  {title} {\bibinfo {title} {Multifractal detrended fluctuation analysis of
  nonstationary time series},\ }\href
  {https://doi.org/10.1016/S0378-4371(02)01383-3} {\bibfield  {journal}
  {\bibinfo  {journal} {Physica A: Statistical Mechanics and its Applications}\
  }\textbf {\bibinfo {volume} {316}},\ \bibinfo {pages} {87} (\bibinfo {year}
  {2002})}\BibitemShut {NoStop}%
\bibitem [{\citenamefont {Hurst}(1951)}]{Hurst1951}%
  \BibitemOpen
  \bibfield  {author} {\bibinfo {author} {\bibfnamefont {H.~E.}\ \bibnamefont
  {Hurst}},\ }\bibfield  {title} {\bibinfo {title} {Long-term storage capacity
  of reservoirs},\ }\href {https://doi.org/10.1061/TACEAT.0006518} {\bibfield
  {journal} {\bibinfo  {journal} {Transactions of the American Society of Civil
  Engineers}\ }\textbf {\bibinfo {volume} {116}},\ \bibinfo {pages} {770}
  (\bibinfo {year} {1951})}\BibitemShut {NoStop}%
\bibitem [{\citenamefont {Simonsen}(2003)}]{Simonsen2003}%
  \BibitemOpen
  \bibfield  {author} {\bibinfo {author} {\bibfnamefont {I.}~\bibnamefont
  {Simonsen}},\ }\bibfield  {title} {\bibinfo {title} {Measuring
  anti-correlations in the nordic electricity spot market by wavelets},\ }\href
  {https://doi.org/10.1016/S0378-4371(02)01938-6} {\bibfield  {journal}
  {\bibinfo  {journal} {Physica A: Statistical Mechanics and its Applications}\
  }\textbf {\bibinfo {volume} {322}},\ \bibinfo {pages} {597} (\bibinfo {year}
  {2003})}\BibitemShut {NoStop}%
\bibitem [{\citenamefont {{\v{C}}urpek}(2019)}]{Curpek2019}%
  \BibitemOpen
  \bibfield  {author} {\bibinfo {author} {\bibfnamefont {J.}~\bibnamefont
  {{\v{C}}urpek}},\ }\bibfield  {title} {\bibinfo {title} {{Time Evolution of
  Hurst Exponent: Czech Wholesale Electricity Market Study}},\ }\href
  {https://doi.org/10.18267/j.efaj.232} {\bibfield  {journal} {\bibinfo
  {journal} {European Financial and Accounting Journal}\ }\textbf {\bibinfo
  {volume} {2019}},\ \bibinfo {pages} {25} (\bibinfo {year}
  {2019})}\BibitemShut {NoStop}%
\bibitem [{\citenamefont {{v}an Kampen}(2007)}]{vanKampen2007}%
  \BibitemOpen
  \bibfield  {author} {\bibinfo {author} {\bibfnamefont {N.~G.}\ \bibnamefont
  {{v}an Kampen}},\ }\href {https://doi.org/10.1016/B978-0-444-52965-7.X5000-4}
  {\emph {\bibinfo {title} {Stochastic Processes in Physics and Chemistry}}},\
  \bibinfo {edition} {3rd}\ ed.\ (\bibinfo  {publisher} {North Holland},\
  \bibinfo {year} {2007})\ \bibinfo {note} {{ISBN:}
  978-0-444-52965-7}\BibitemShut {NoStop}%
\bibitem [{EPE(2021{\natexlab{b}})}]{EPEX_record}%
  \BibitemOpen
  \href@noop {} {\bibinfo {title} {{New record volume traded on EPEX SPOT in
  2020}}} (\bibinfo {year} {2021}{\natexlab{b}}),\ \bibinfo {note} {14 January
  2021,
  \href{https://www.epexspot.com/en/news/new-record-volume-traded-epex-spot-2020}{https://www.epexspot.com/en/news/new-record-volume-traded-epex-spot-2020}}\BibitemShut
  {NoStop}%
\bibitem [{\citenamefont {Märkle-Hu\ss}\ \emph {et~al.}(2018)\citenamefont
  {Märkle-Hu\ss}, \citenamefont {Feuerriegel},\ and\ \citenamefont
  {Neumann}}]{MaerkleHuss2018}%
  \BibitemOpen
  \bibfield  {author} {\bibinfo {author} {\bibfnamefont {J.}~\bibnamefont
  {Märkle-Hu\ss}}, \bibinfo {author} {\bibfnamefont {S.}~\bibnamefont
  {Feuerriegel}},\ and\ \bibinfo {author} {\bibfnamefont {D.}~\bibnamefont
  {Neumann}},\ }\bibfield  {title} {\bibinfo {title} {Contract durations in the
  electricity market: {C}ausal impact of 15 min trading on the {EPEX SPOT}
  market},\ }\href {https://doi.org/10.1016/j.eneco.2017.11.019} {\bibfield
  {journal} {\bibinfo  {journal} {Energy Economics}\ }\textbf {\bibinfo
  {volume} {69}},\ \bibinfo {pages} {367} (\bibinfo {year} {2018})}\BibitemShut
  {NoStop}%
\bibitem [{EPE(2011)}]{EPEX_15min}%
  \BibitemOpen
  \href@noop {} {\bibinfo {title} {{Press Release: 15-minute contracts
  successfully launched on German Intraday market}}} (\bibinfo {year} {2011}),\
  \bibinfo {note} {15 December 2011,
  \href{https://www.epexspot.com/sites/default/files/download_center_files/2011-12-15_15\%20minute_contracts_launch.pdf}{https://www.epexspot.com/sites/default/files/download
  \_center\_files/2011-12-15\_15\%20minute\_contracts
  \_launch.pdf}}\BibitemShut {NoStop}%
\bibitem [{\citenamefont {Kruse}\ \emph
  {et~al.}(2021{\natexlab{a}})\citenamefont {Kruse}, \citenamefont
  {Sch{\"a}fer},\ and\ \citenamefont {Witthaut}}]{Kruse2021a}%
  \BibitemOpen
  \bibfield  {author} {\bibinfo {author} {\bibfnamefont {J.}~\bibnamefont
  {Kruse}}, \bibinfo {author} {\bibfnamefont {B.}~\bibnamefont {Sch{\"a}fer}},\
  and\ \bibinfo {author} {\bibfnamefont {D.}~\bibnamefont {Witthaut}},\
  }\bibfield  {title} {\bibinfo {title} {Revealing drivers and risks for power
  grid frequency stability with explainable {AI}},\ }\href@noop {} {\bibfield
  {journal} {\bibinfo  {journal}
  {\href{https://arxiv.org/abs/2106.04341}{preprint arXiv:2106.04341}}\ }
  (\bibinfo {year} {2021}{\natexlab{a}})}\BibitemShut {NoStop}%
\bibitem [{\citenamefont {Kruse}\ \emph
  {et~al.}(2021{\natexlab{b}})\citenamefont {Kruse}, \citenamefont
  {Sch{\"a}fer},\ and\ \citenamefont {Witthaut}}]{Kruse2021b}%
  \BibitemOpen
  \bibfield  {author} {\bibinfo {author} {\bibfnamefont {J.}~\bibnamefont
  {Kruse}}, \bibinfo {author} {\bibfnamefont {B.}~\bibnamefont {Sch{\"a}fer}},\
  and\ \bibinfo {author} {\bibfnamefont {D.}~\bibnamefont {Witthaut}},\
  }\bibfield  {title} {\bibinfo {title} {Secondary control activation analysed
  and predicted with explainable {AI}},\ }\href@noop {} {\bibfield  {journal}
  {\bibinfo  {journal} {\textit{in preparation}}\ } (\bibinfo {year}
  {2021}{\natexlab{b}})}\BibitemShut {NoStop}%
\bibitem [{\citenamefont {Ocker}\ and\ \citenamefont
  {Ehrhart}(2017)}]{Ocker2017}%
  \BibitemOpen
  \bibfield  {author} {\bibinfo {author} {\bibfnamefont {F.}~\bibnamefont
  {Ocker}}\ and\ \bibinfo {author} {\bibfnamefont {K.-M.}\ \bibnamefont
  {Ehrhart}},\ }\bibfield  {title} {\bibinfo {title} {The ``{G}erman
  {P}aradox'' in the balancing power markets},\ }\href
  {https://doi.org/10.1016/j.rser.2016.09.040} {\bibfield  {journal} {\bibinfo
  {journal} {Renewable and Sustainable Energy Reviews}\ }\textbf {\bibinfo
  {volume} {67}},\ \bibinfo {pages} {892} (\bibinfo {year} {2017})}\BibitemShut
  {NoStop}%
\bibitem [{\citenamefont {Koscielny-Bunde}\ \emph {et~al.}(1998)\citenamefont
  {Koscielny-Bunde}, \citenamefont {Bunde}, \citenamefont {Havlin},
  \citenamefont {Roman}, \citenamefont {Goldreich},\ and\ \citenamefont
  {Schellnhuber}}]{KoscielnyBunde1998}%
  \BibitemOpen
  \bibfield  {author} {\bibinfo {author} {\bibfnamefont {E.}~\bibnamefont
  {Koscielny-Bunde}}, \bibinfo {author} {\bibfnamefont {A.}~\bibnamefont
  {Bunde}}, \bibinfo {author} {\bibfnamefont {S.}~\bibnamefont {Havlin}},
  \bibinfo {author} {\bibfnamefont {H.~E.}\ \bibnamefont {Roman}}, \bibinfo
  {author} {\bibfnamefont {Y.}~\bibnamefont {Goldreich}},\ and\ \bibinfo
  {author} {\bibfnamefont {H.-J.}\ \bibnamefont {Schellnhuber}},\ }\bibfield
  {title} {\bibinfo {title} {Indication of a universal persistence law
  governing atmospheric variability},\ }\href
  {https://doi.org/10.1103/PhysRevLett.81.729} {\bibfield  {journal} {\bibinfo
  {journal} {Physical Review Letters}\ }\textbf {\bibinfo {volume} {81}},\
  \bibinfo {pages} {729} (\bibinfo {year} {1998})}\BibitemShut {NoStop}%
\bibitem [{\citenamefont {Tabar}(2019)}]{Tabar2019}%
  \BibitemOpen
  \bibfield  {author} {\bibinfo {author} {\bibfnamefont {M.~R.~R.}\
  \bibnamefont {Tabar}},\ }\href {https://doi.org/10.1007/978-3-030-18472-8}
  {\emph {\bibinfo {title} {Analysis and Data-Based Reconstruction of Complex
  Nonlinear Dynamical Systems}}},\ \bibinfo {edition} {1st}\ ed.\ (\bibinfo
  {publisher} {Springer International Publishing},\ \bibinfo {year} {2019})\
  \bibinfo {note} {{ISBN}: 978-3-030-18471-1}\BibitemShut {NoStop}%
\bibitem [{\citenamefont {Ihlen}(2012)}]{Ihlen2012}%
  \BibitemOpen
  \bibfield  {author} {\bibinfo {author} {\bibfnamefont {E.}~\bibnamefont
  {Ihlen}},\ }\bibfield  {title} {\bibinfo {title} {Introduction to
  {M}ultifractal {D}etrended {F}luctuation {A}nalysis in {M}atlab},\ }\href
  {https://doi.org/10.3389/fphys.2012.00141} {\bibfield  {journal} {\bibinfo
  {journal} {Frontiers in Physiology}\ }\textbf {\bibinfo {volume} {3}},\
  \bibinfo {pages} {141} (\bibinfo {year} {2012})}\BibitemShut {NoStop}%
\bibitem [{\citenamefont {Rydin Gorj\~ao}\ \emph {et~al.}(2021)\citenamefont
  {Rydin Gorj\~ao}, \citenamefont {Hassan}, \citenamefont {Kurths},\ and\
  \citenamefont {Witthaut}}]{RydinGorjao2021}%
  \BibitemOpen
  \bibfield  {author} {\bibinfo {author} {\bibfnamefont {L.}~\bibnamefont
  {Rydin Gorj\~ao}}, \bibinfo {author} {\bibfnamefont {G.}~\bibnamefont
  {Hassan}}, \bibinfo {author} {\bibfnamefont {J.}~\bibnamefont {Kurths}},\
  and\ \bibinfo {author} {\bibfnamefont {D.}~\bibnamefont {Witthaut}},\
  }\bibfield  {title} {\bibinfo {title} {{\texttt{MFDFA}}: Efficient
  multifractal detrended fluctuation analysis in python},\ }\href@noop {}
  {\bibfield  {journal} {\bibinfo  {journal}
  {\href{https://arxiv.org/abs/2104.10470}{preprint arXiv:2104.10470}}\ }
  (\bibinfo {year} {2021})}\BibitemShut {NoStop}%
\bibitem [{\citenamefont {Peng}\ \emph {et~al.}(1995)\citenamefont {Peng},
  \citenamefont {Havlin}, \citenamefont {Stanley},\ and\ \citenamefont
  {Goldberger}}]{Peng1995}%
  \BibitemOpen
  \bibfield  {author} {\bibinfo {author} {\bibfnamefont {C.}~\bibnamefont
  {Peng}}, \bibinfo {author} {\bibfnamefont {S.}~\bibnamefont {Havlin}},
  \bibinfo {author} {\bibfnamefont {H.~E.}\ \bibnamefont {Stanley}},\ and\
  \bibinfo {author} {\bibfnamefont {A.~L.}\ \bibnamefont {Goldberger}},\
  }\bibfield  {title} {\bibinfo {title} {Quantification of scaling exponents
  and crossover phenomena in nonstationary heartbeat time series},\ }\href
  {https://doi.org/10.1063/1.166141} {\bibfield  {journal} {\bibinfo  {journal}
  {Chaos: An Interdisciplinary Journal of Nonlinear Science}\ }\textbf
  {\bibinfo {volume} {5}},\ \bibinfo {pages} {82} (\bibinfo {year}
  {1995})}\BibitemShut {NoStop}%
\bibitem [{\citenamefont {Kramers}(1940)}]{Kramers1940}%
  \BibitemOpen
  \bibfield  {author} {\bibinfo {author} {\bibfnamefont {H.~A.}\ \bibnamefont
  {Kramers}},\ }\bibfield  {title} {\bibinfo {title} {Brownian motion in a
  field of force and the diffusion model of chemical reactions},\ }\href
  {https://doi.org/10.1016/S0031-8914(40)90098-2} {\bibfield  {journal}
  {\bibinfo  {journal} {Physica}\ }\textbf {\bibinfo {volume} {7}},\ \bibinfo
  {pages} {284} (\bibinfo {year} {1940})}\BibitemShut {NoStop}%
\bibitem [{\citenamefont {Moyal}(1949)}]{Moyal1949}%
  \BibitemOpen
  \bibfield  {author} {\bibinfo {author} {\bibfnamefont {J.~E.}\ \bibnamefont
  {Moyal}},\ }\bibfield  {title} {\bibinfo {title} {Stochastic processes and
  statistical physics},\ }\href@noop {} {\bibfield  {journal} {\bibinfo
  {journal} {Journal of the Royal Statistical Society: Series B
  (Methodological)}\ }\textbf {\bibinfo {volume} {11}},\ \bibinfo {pages} {150}
  (\bibinfo {year} {1949})},\ \bibinfo {note}
  {\href{http://www.jstor.org/stable/2984076}{http://www.jstor.org/stable/2984076}}\BibitemShut
  {NoStop}%
\bibitem [{\citenamefont {Friedrich}\ and\ \citenamefont
  {Peinke}(1997)}]{Friedrich1997}%
  \BibitemOpen
  \bibfield  {author} {\bibinfo {author} {\bibfnamefont {R.}~\bibnamefont
  {Friedrich}}\ and\ \bibinfo {author} {\bibfnamefont {J.}~\bibnamefont
  {Peinke}},\ }\bibfield  {title} {\bibinfo {title} {Description of a turbulent
  cascade by a {F}okker-{P}lanck equation},\ }\href
  {https://doi.org/10.1103/PhysRevLett.78.863} {\bibfield  {journal} {\bibinfo
  {journal} {Physical Review Letters}\ }\textbf {\bibinfo {volume} {78}},\
  \bibinfo {pages} {863} (\bibinfo {year} {1997})}\BibitemShut {NoStop}%
\bibitem [{\citenamefont {Müller}\ \emph {et~al.}(2018)\citenamefont
  {Müller}, \citenamefont {Plath}, \citenamefont {Radons},\ and\ \citenamefont
  {Fuchs}}]{Mueller2018}%
  \BibitemOpen
  \bibinfo {editor} {\bibfnamefont {S.~C.}\ \bibnamefont {Müller}}, \bibinfo
  {editor} {\bibfnamefont {P.~J.}\ \bibnamefont {Plath}}, \bibinfo {editor}
  {\bibfnamefont {G.}~\bibnamefont {Radons}},\ and\ \bibinfo {editor}
  {\bibfnamefont {A.}~\bibnamefont {Fuchs}},\ eds.,\ \href
  {https://doi.org/10.1007/978-3-319-64334-2} {\emph {\bibinfo {title}
  {Complexity and Synergetics}}},\ \bibinfo {edition} {1st}\ ed.\ (\bibinfo
  {publisher} {Springer International Publishing},\ \bibinfo {year} {2018})\
  \bibinfo {note} {{ISBN}: 978-3-319-64333-5}\BibitemShut {NoStop}%
\bibitem [{\citenamefont {Friedrich}\ \emph {et~al.}(2018)\citenamefont
  {Friedrich}, \citenamefont {Margazoglou}, \citenamefont {Biferale},\ and\
  \citenamefont {Grauer}}]{Friedrich2018}%
  \BibitemOpen
  \bibfield  {author} {\bibinfo {author} {\bibfnamefont {J.}~\bibnamefont
  {Friedrich}}, \bibinfo {author} {\bibfnamefont {G.}~\bibnamefont
  {Margazoglou}}, \bibinfo {author} {\bibfnamefont {L.}~\bibnamefont
  {Biferale}},\ and\ \bibinfo {author} {\bibfnamefont {R.}~\bibnamefont
  {Grauer}},\ }\bibfield  {title} {\bibinfo {title} {Multiscale velocity
  correlations in turbulence and {B}urgers turbulence: {F}usion rules, {M}arkov
  processes in scale, and multifractal predictions},\ }\href
  {https://doi.org/10.1103/PhysRevE.98.023104} {\bibfield  {journal} {\bibinfo
  {journal} {Physical Review E}\ }\textbf {\bibinfo {volume} {98}},\ \bibinfo
  {pages} {023104} (\bibinfo {year} {2018})}\BibitemShut {NoStop}%
\bibitem [{\citenamefont {Friedrich}\ and\ \citenamefont
  {Grauer}(2020)}]{Friedrich2020}%
  \BibitemOpen
  \bibfield  {author} {\bibinfo {author} {\bibfnamefont {J.}~\bibnamefont
  {Friedrich}}\ and\ \bibinfo {author} {\bibfnamefont {R.}~\bibnamefont
  {Grauer}},\ }\bibfield  {title} {\bibinfo {title} {Generalized description of
  intermittency in turbulence via stochastic methods},\ }\href
  {https://doi.org/10.3390/atmos11091003} {\bibfield  {journal} {\bibinfo
  {journal} {Atmosphere}\ }\textbf {\bibinfo {volume} {11}},\ \bibinfo {pages}
  {1003} (\bibinfo {year} {2020})}\BibitemShut {NoStop}%
\bibitem [{\citenamefont {Nickelsen}(2017)}]{Nickelsen2017}%
  \BibitemOpen
  \bibfield  {author} {\bibinfo {author} {\bibfnamefont {D.}~\bibnamefont
  {Nickelsen}},\ }\bibfield  {title} {\bibinfo {title} {Master equation for
  {S}he{\textendash}{L}eveque scaling and its classification in terms of other
  {M}arkov models of developed turbulence},\ }\href
  {https://doi.org/10.1088/1742-5468/aa786a} {\bibfield  {journal} {\bibinfo
  {journal} {Journal of Statistical Mechanics: Theory and Experiment}\ }\textbf
  {\bibinfo {volume} {2017}},\ \bibinfo {pages} {073209} (\bibinfo {year}
  {2017})}\BibitemShut {NoStop}%
\bibitem [{\citenamefont {Rydin Gorj\~ao}\ and\ \citenamefont
  {Meirinhos}(2019)}]{RydinGorjao2019}%
  \BibitemOpen
  \bibfield  {author} {\bibinfo {author} {\bibfnamefont {L.}~\bibnamefont
  {Rydin Gorj\~ao}}\ and\ \bibinfo {author} {\bibfnamefont {F.}~\bibnamefont
  {Meirinhos}},\ }\bibfield  {title} {\bibinfo {title} {\texttt{kramersmoyal}:
  {K}ramers--{M}oyal coefficients for stochastic processes},\ }\href
  {https://doi.org/10.21105/joss.01693} {\bibfield  {journal} {\bibinfo
  {journal} {Journal of Open Source Software}\ }\textbf {\bibinfo {volume}
  {4}},\ \bibinfo {pages} {1693} (\bibinfo {year} {2019})}\BibitemShut
  {NoStop}%
\bibitem [{\citenamefont {Lamouroux}\ and\ \citenamefont
  {Lehnertz}(2009)}]{Lamouroux2009}%
  \BibitemOpen
  \bibfield  {author} {\bibinfo {author} {\bibfnamefont {D.}~\bibnamefont
  {Lamouroux}}\ and\ \bibinfo {author} {\bibfnamefont {K.}~\bibnamefont
  {Lehnertz}},\ }\bibfield  {title} {\bibinfo {title} {Kernel-based regression
  of drift and diffusion coefficients of stochastic processes},\ }\href
  {https://doi.org/10.1016/j.physleta.2009.07.073} {\bibfield  {journal}
  {\bibinfo  {journal} {Physics Letters A}\ }\textbf {\bibinfo {volume}
  {373}},\ \bibinfo {pages} {3507} (\bibinfo {year} {2009})}\BibitemShut
  {NoStop}%
\bibitem [{\citenamefont {Rydin~Gorjão}\ \emph {et~al.}(2021)\citenamefont
  {Rydin~Gorjão}, \citenamefont {Witthaut}, \citenamefont {Lehnertz},\ and\
  \citenamefont {Lind}}]{RydinGorjao2021b}%
  \BibitemOpen
  \bibfield  {author} {\bibinfo {author} {\bibfnamefont {L.}~\bibnamefont
  {Rydin~Gorjão}}, \bibinfo {author} {\bibfnamefont {D.}~\bibnamefont
  {Witthaut}}, \bibinfo {author} {\bibfnamefont {K.}~\bibnamefont {Lehnertz}},\
  and\ \bibinfo {author} {\bibfnamefont {P.~G.}\ \bibnamefont {Lind}},\
  }\bibfield  {title} {\bibinfo {title} {Arbitrary-order finite-time
  corrections for the {K}ramers--{M}oyal operator},\ }\href
  {https://doi.org/10.3390/e23050517} {\bibfield  {journal} {\bibinfo
  {journal} {Entropy}\ }\textbf {\bibinfo {volume} {23}},\ \bibinfo {pages}
  {517} (\bibinfo {year} {2021})}\BibitemShut {NoStop}%
\bibitem [{\citenamefont {Sadegh~Movahed}\ \emph {et~al.}(2006)\citenamefont
  {Sadegh~Movahed}, \citenamefont {Jafari}, \citenamefont {Ghasemi},
  \citenamefont {Rahvar},\ and\ \citenamefont {Tabar}}]{Movahed2006}%
  \BibitemOpen
  \bibfield  {author} {\bibinfo {author} {\bibfnamefont {M.}~\bibnamefont
  {Sadegh~Movahed}}, \bibinfo {author} {\bibfnamefont {G.~R.}\ \bibnamefont
  {Jafari}}, \bibinfo {author} {\bibfnamefont {F.}~\bibnamefont {Ghasemi}},
  \bibinfo {author} {\bibfnamefont {S.}~\bibnamefont {Rahvar}},\ and\ \bibinfo
  {author} {\bibfnamefont {M.~R.~R.}\ \bibnamefont {Tabar}},\ }\bibfield
  {title} {\bibinfo {title} {Multifractal detrended fluctuation analysis of
  sunspot time series},\ }\href
  {https://doi.org/10.1088/1742-5468/2006/02/p02003} {\bibfield  {journal}
  {\bibinfo  {journal} {Journal of Statistical Mechanics: Theory and
  Experiment}\ }\textbf {\bibinfo {volume} {2006}},\ \bibinfo {pages} {P02003}
  (\bibinfo {year} {2006})}\BibitemShut {NoStop}%
\end{thebibliography}%

\end{document}